\documentclass[journal,10pt]{IEEEtran}

\def\ps@headings{%
	\def\@oddhead{\mbox{}\scriptsize\rightmark \hfil \thepage}%
	\def\@evenhead{\scriptsize\thepage \hfil \leftmark\mbox{}}%
	\def\@oddfoot{}%
	\def\@evenfoot{}}
\makeatother

\IEEEoverridecommandlockouts

\usepackage{cite}
\usepackage{amsmath,amssymb,amsfonts}
\usepackage{algorithmic}
\usepackage{textcomp}
\usepackage{bm}
\usepackage[font=small]{caption}
\usepackage{subfig}
\usepackage{graphicx}
\usepackage{graphics}
\usepackage{epstopdf}
\usepackage{xcolor}
\usepackage{dblfloatfix}    % To enable figures at the bottom of page
\usepackage{verbatim}

\usepackage{tabulary}
\usepackage{lipsum}
\usepackage{hyperref}
\usepackage{enumitem}
\usepackage{multicol}
\usepackage{gensymb}
\usepackage{mathrsfs}
\usepackage{booktabs}% http://ctan.org/pkg/booktabs
\usepackage[many]{tcolorbox}
\usepackage{soul}

\def\BibTeX{{\rm B\kern-.05em{\sc i\kern-.025em b}\kern-.08em
		T\kern-.1667em\lower.7ex\hbox{E}\kern-.125emX}}

\newcommand*{\BBE}{\mathbb{E}}
\makeatletter
\newcommand*{\rom}[1]{\expandafter\@slowromancap\romannumeral #1@}
\makeatother
\DeclareMathOperator{\Tr}{Tr}
\renewcommand{\theenumi}{label=\Alph{enumi}}

\IEEEoverridecommandlockouts

\begin{document}
\pagenumbering{gobble}
\title{5G Massive MIMO Architectures: Self-Backhauled Small Cells versus Direct Access}

\author{\IEEEauthorblockN{Andrea~Bonfante, Lorenzo~Galati~Giordano, David~L\'{o}pez-P\'{e}rez, Adrian~Garcia-Rodriguez, \\ Giovanni~Geraci, Paolo~Baracca, M.~Majid~Butt, and~Nicola~Marchetti}%
\thanks{This work was supported in part by Irish Research Council, by Nokia Ireland Ltd under Grant Number EPSPG/2016/106, and by Science Foundation Ireland (SFI) under the European Regional Development Fund -- Grant Number 13/RC/2077.}%
\thanks{A. Bonfante is with Nokia Bell Labs, Dublin, Ireland and with CONNECT Centre for Future Networks, Trinity College Dublin, Ireland (e-mail: andrea.bonfante@nokia-bell-labs.com).}%
\thanks{L. Galati Giordano, D. L\'{o}pez-P\'{e}rez, and A. Garcia-Rodriguez are with Nokia Bell Labs, Dublin, Ireland (e-mails: lorenzo.galati\_giordano@nokia-bell-labs.com; david.lopez-perez@nokia-bell-labs.com; adrian.garcia\_rodriguez@nokia-bell-labs.com).}%
\thanks{G. Geraci was with Nokia Bell Labs, Dublin, Ireland. He is now with the Department of Information and Communication Technologies, Universitat Pompeu Fabra, Barcelona, Spain (e-mail: dr.giovanni.geraci@gmail.com).}%
\thanks{P. Baracca is with Nokia Bell Labs, Stuttgart, Germany (e-mail: paolo.baracca@nokia-bell-labs.com).}%
\thanks{M. M. Butt was with University of Glasgow, Glasgow, U.K. He is now with Nokia Bell Labs, Paris--Saclay, France (e-mail: majid.butt@nokia-bell-labs.com).}%
\thanks{N. Marchetti is with CONNECT Centre for Future Networks, Trinity College Dublin, Ireland (e-mail: nicola.marchetti@tcd.ie).}
\thanks{A part of this paper was presented at IEEE Globecom 2018 \cite{8647638}.}}%

\maketitle

\thispagestyle{empty}

% Abstract
\begin{abstract}
    In this paper, 
	we focus on one of the key technologies for the fifth-generation wireless communication networks, 
	massive multiple-input-multiple-output (mMIMO), 
	by investigating two of its most relevant architectures: 
	\emph{1)} to provide in-band backhaul for the ultra-dense network (UDN) of self-backhauled small cells (SCs),
	and \emph{2)} to provide direct access (DA) to user equipments (UEs). 
	Through comprehensive 3GPP-based system-level simulations and analytical formulations, 
	we show the end-to-end UE rates achievable with these two architectures.
	Differently from the existing works, we provide results for two strategies of self-backhauled SC deployments, namely \emph{random} and \emph{ad-hoc}, where in the latter SCs are purposely positioned close to UEs to achieve line-of-sight (LoS) access links.
	We also evaluate the optimal backhaul and access time resource partition due to the in-band self-backhauling (s-BH) operations. 
	Our results show that the ad-hoc deployment of self-backhauled SCs closer to the UEs with optimal resource partition and with directive antenna patterns,
	provides rate improvements for cell-edge UEs that amount to $30\%$ and tenfold gain, as compared to mMIMO DA architecture with pilot reuse 3 and reuse 1, respectively.
	On the other hand, 
	mMIMO s-BH underperforms mMIMO DA above the median value of the UE rates when the effect of pilot contamination is less severe, and the LoS probability of the DA links improves.
\end{abstract}

% Keywords

\IEEEpeerreviewmaketitle
% Make space for copyright footnote
%\IEEEpubidadjcol
% Column break 
%\vfill\eject

\section{Introduction}

\IEEEpubidadjcol

\IEEEPARstart{F}{ifth}-generation (5G) wireless communication systems are expected to support a 1000x increase in capacity compared to existing networks \cite{7169508}. 
Meeting this gargantuan target will require mobile network operators (MNOs) to leverage new technologies, 
such as massive multiple-input-multiple-output (mMIMO), 
and network densification using small cells base stations (SCs) \cite{6375940, 7126919}.
More specifically, mMIMO consists of deploying a large number of antennas at the base station (BS) to enable the simultaneous transmission to and from a multitude of terminals which are spatially multiplexed in the same time-frequency resources \cite{6375940}, while network densification considers the deployment of a large number of SCs up to the point of having one SC per user equipment (UE) \cite{7126919}.
A further step towards the practical ultra-dense SC deployment is represented by self-backhauling (s-BH), defined as the capability for a BS to enable in the same spectrum resources both the access and the backhaul transmissions.
Self-backhauling could drive the tight integration between mMIMO and ultra-dense SC deployment luring MNOs with the potential of achieving the desired capacity boost at a contained investment \cite{7306534}.
Indeed, exploiting the large number of spatial degrees-of-freedom (DoF) available with mMIMO to provide sub-6\,GHz in-band wireless backhauling to SCs offers multiple advantages to MNOs: 
\emph{i)} avoiding deployment of an expensive wired backhaul infrastructure, 
\emph{ii)} availing of more flexibility in the deployment of SCs, 
and \emph{iii)} not having to purchase additional licensed spectrum, 
as in the case of out-of-band wireless backhauling \cite{7306536}.

Those advantages motivated the Third Generation Partnership Project (3GPP) to include in the 5G New Radio (NR) Release 15 a new study item, 
which focuses on Integrated Access and Backhaul (IAB) network architectures -- also refereed to as \emph{self-backhauling} networks in the literature.
In \cite{TR_38.874}, 
3GPP provides a list of use cases, 
in both the sub-6 GHz and above-6 GHz spectrum bands,
and network architecture requirements for the NR backhauling functionalities coupled with the radio access network (RAN) technology.
%The envisioned scenarios for IAB comprise the support of 
%NR access traffic over NR backhaul links and legacy Long Term Evolution (LTE) access traffic over NR backhaul links.
%Moreover, IAB can support access and backhaul for both sub-6 GHz and above-6 GHz spectrum for in-band and out-of-band scenarios. 

\subsection{Background and Related Work}

% needed in second column of first page if using \IEEEpubid
\IEEEpubidadjcol

Several works focused on millimeter wave (mmWave) s-BH networks \cite{7974805,Singh2015}, %\cite{7974805,Singh2015,2017arXiv171006255S}
which offer wide bandwidth channels to accommodate multiple backhaul and access links simultaneously.
At the same time, various research efforts considered sub-6 GHz s-BH networks in a heterogeneous network (HetNet) environment \cite{7962657,Nguyen2016,7817893}, 
which is more suitable to provide wide-area coverage through conventional macro-cells, 
and use self-backhauled SCs to further boost the network capacity.
However, due to the scarcity of spectrum in the bands below 6 GHz, 
the bandwidth splitting required to serve multiple backhaul links and the inter-tier interference caused by co-channel access and backhaul operations may turn out to be a significant impediment to the potential adoption of s-BH in sub-6 GHz HetNets \cite{7817893}. 
In \cite{7962657,Nguyen2016}, the authors tackled the problem of resource allocation (such as transmission power and time-frequency resources) of s-BH networks. 
In \cite{7817893}, the authors considered a full-duplex (FD) technique, which on the one hand avoids the partitioning of the spectrum between backhaul and access, but on the other hand, it requires to study the problem of the self-interference due to the possible bi-directional transmission.

Moreover, works such as \cite{7177124,Tabassum}, considered macro BSs equipped with mMIMO to enhance the backhaul link capacity and simultaneously serve SCs and UEs.
In \cite{7177124}, the authors studied the UE data-rate performance as a function of the distance between the mMIMO-BS and the s-BH SCs. However, they considered a simplified single-cell scenario without inter-cell interference between SCs.
In \cite{Tabassum,7445888_GioBackhaul}, the authors used stochastic geometry to derive the rate coverage probability and compute the optimal proportion of in-band and out-of-band FD SCs in the network which maximizes the UE rates, and energy efficiency, respectively.
Finally, the authors in \cite{8241817} investigated an optimization approach to maximize the sum-rate of the UEs under capacity constrained backhaul and by considering the length of the pilot sequences used for the channel state information (CSI) acquisition.

%Furthermore, works such as \cite{7511179,8053839} carried out the comparison between mMIMO and dense SC networks on the basic assumption that the SCs are equipped with a wired backhaul. 

\subsection{Motivation and Contribution}

In this paper, we analyze the end-to-end UE performance of mMIMO s-BH architecture by means of theoretical analysis and 3GPP-based system-level simulations when compared to mMIMO Direct Access (DA),
where mMIMO-BSs are solely dedicated to serving UEs in the absence of SCs \cite{Lim}. 
We consider a realistic multi-cell setup \cite{3gpp.36.814}, 
where mMIMO-BSs provide sub-6\,GHz backhauling to a plurality of half-duplex (HD) SCs overlaying the macro cellular area.
In these HD systems, a s-BH network entails sharing time-and-frequency resources between radio access and backhaul links.
We use 3GPP-based system level simulations up to the point of one SC deployed for each UE, whereas we use the analytical framework when the SC density is much larger than the UE density. In this regime, numerical simulations become impractical due to the large amount of time required and the high computational complexity involved.
Two different strategies of self-backhauled SC deployments are considered as illustrated in Fig. \ref{fig:network_layout}. 
We analyze a \emph{random} deployment -- where SCs are uniformly distributed over a geographical area --, 
and an \emph{ad-hoc} deployment -- where SCs are purposely positioned close to UEs to achieve line-of-sight (LoS) access links. 
We would like to stress that the particular choice of the ad-hoc SC deployment serves two purposes: \emph{a)} to study the performance gains that can be obtained by the MNOs through a tailored SCs deployment, and \emph{b)} to provide an upper bound of the system access performance obtained when the SC locations are coupled to the UE locations.

%Indeed, the latter type of deployment can be supported by future dynamic network infrastructures, for example based on the applications of unmanned aerial vehicles (UAV) to carry SCs \cite{7744808}. 

The contributions of the paper are as follows:
\begin{enumerate}
	
	\item[1)] 
		
	We provide 3GPP-based system-level simulations results on the performance of the achievable UE data-rates in mMIMO based wireless in-band s-BH with random and ad-hoc SC deployments.
	%The latter is used as an exercise to give an upper bound on the achievable capacity of the s-BH architecture.
	%To the best of the authors knowledge, we also compare for the first time the performance of mMIMO s-BH and mMIMO DA architectures.
	Differently from the previous works \cite{Tabassum,7445888_GioBackhaul,7177124,8241817}, 
	our work accounts for a pathloss model with LoS and Non-LoS (NLoS) transitions in both backhaul and access links. 
	The closest similar work is \cite{7177124}, 
	which also compares the performance of mMIMO s-BH and mMIMO DA architectures by considering channel models based only on pathloss, 
	but not accounting for the pilot contamination effect on the signal-to-interference-plus-noise ratio (SINR)	when mMIMO is used.
	On the contrary, we show that both the pathloss models incorporating LoS and NLoS transmissions and the pilot contamination effect severely impact the inter-cell interference modeling and the system performance, 
	making our results more accurate and realistic than those presented in \cite{7177124}.
	
	\item[2)]
	
	We provide an analytical model for evaluating the average data-rate of the backhaul and access links.
	We adapt the expressions proposed in \cite{Kelif2010,6702841} to model the mMIMO backhaul links. 
	Specifically, 
	we account for the effects of the antenna directionality and sectorization, 
	and the effect of the beamforming gain due to the mMIMO precoding.	
	In the access link formulation, we account for the density of active SCs, which matches the numerical results obtained by simulations.
	We also employ the analytical framework to show that, due to the over provisioning of self-backhauled SCs, random deployment requires thousands of SCs to achieve the same performance of the ad-hoc deployment.
	
	\item[3)]
		
	We explain in details the different factors playing a key role in the 3GPP-based system-level results.
	Overall, the insights from these results can guide the deployment of future 5G access network.
	
\end{enumerate}

The remainder of this paper is organized as follows.
Section \hyperref[Sec:2]{\rom{2}} introduces the system model on which the analysis is based;
Section \hyperref[Sec:3]{\rom{3}} presents the downlink (DL) SINR and rate expressions of the backhaul and access links; 
Section \hyperref[Sec:4]{\rom{4}} provides the analytical signal-to-interference ratio (SIR) and average rate expressions of backhaul and access; 
Section \hyperref[Sec:5]{\rom{5}} presents the numerical results, 
and \hyperref[Sec:5]{\rom{6}} summarizes the key findings.
The following notation is used throughout the paper.
Capital and lower-case bold letters denote matrices and vectors, respectively, while $[\cdot]^*$, $[\cdot]^\mathrm{T}$ and $[\cdot]^\mathrm{H}$ denote conjugate, transpose, and conjugate transpose, respectively.
$| \cdot |$ indicates the absolute value.
The notation ${\BBE}[\cdot]$ indicates the expectation with respect to a random variable (RV).
We use $\mathbb{C}$ and $\mathbb{R}$ to denote complex- and real-valued numbers, respectively.

\section{System Model}\label{Sec:2}

\begin{figure}
	\centering
	\subfloat[s-BH architecture with random deployment]{\includegraphics[width=0.90\columnwidth]{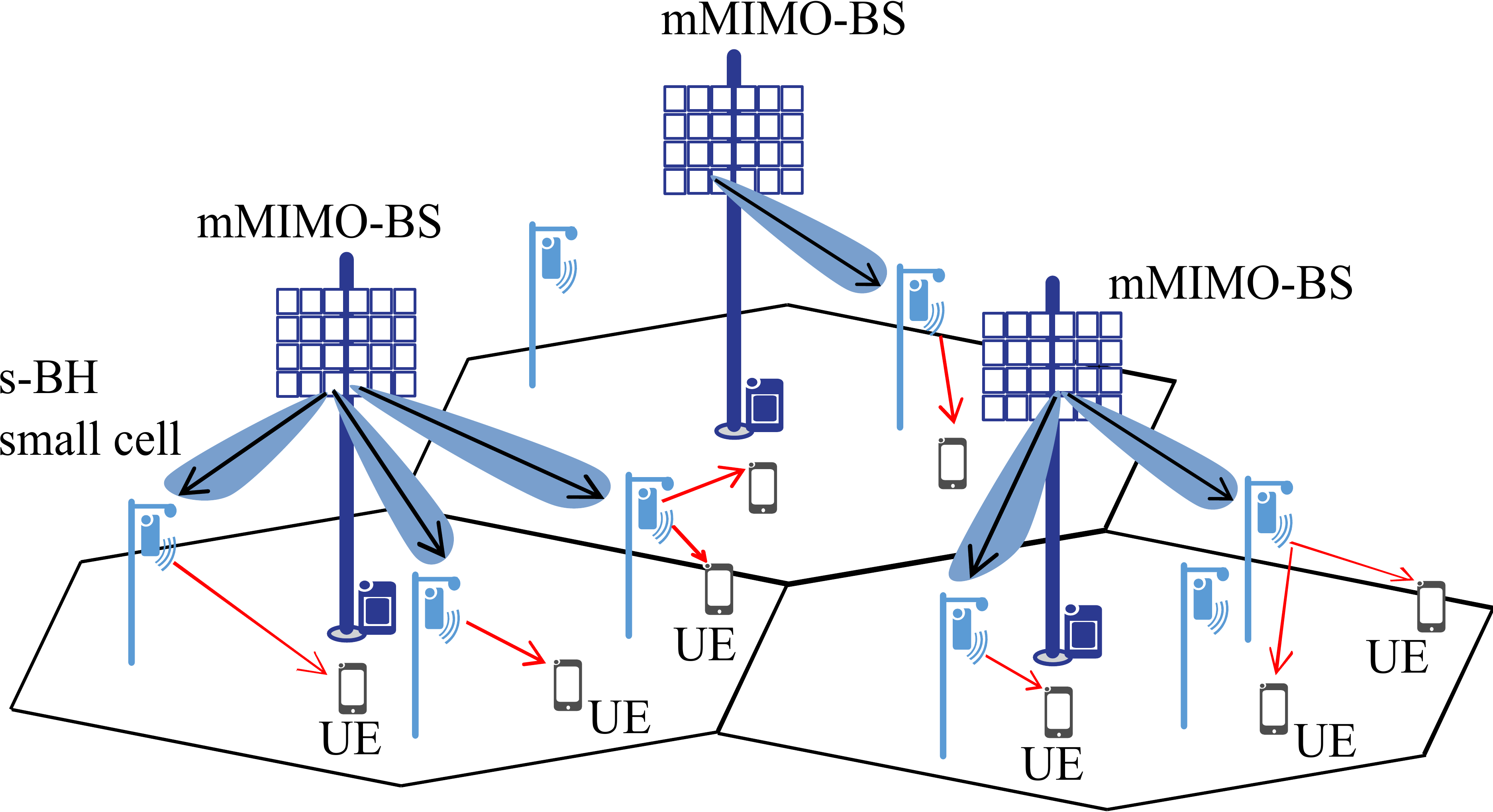}\label{fig1a}}%
	\hfill
	%\hspace{0.2in}
	\subfloat[s-BH architecture with ad-hoc deployment]{\includegraphics[width=0.90\columnwidth]{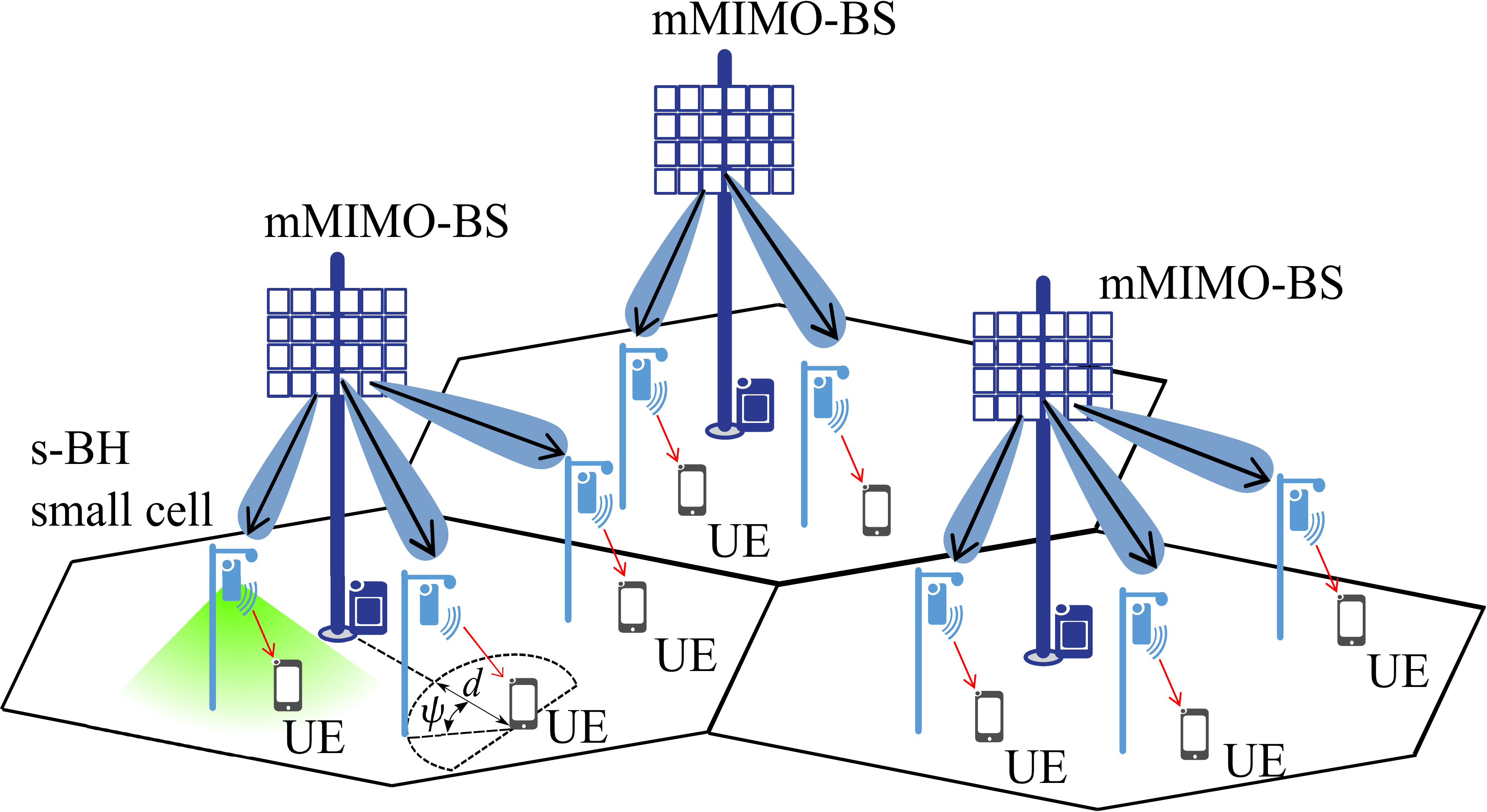}\label{fig1b}}%
	\hfill
	\subfloat[DA architecture]{\includegraphics[width=0.90\columnwidth]{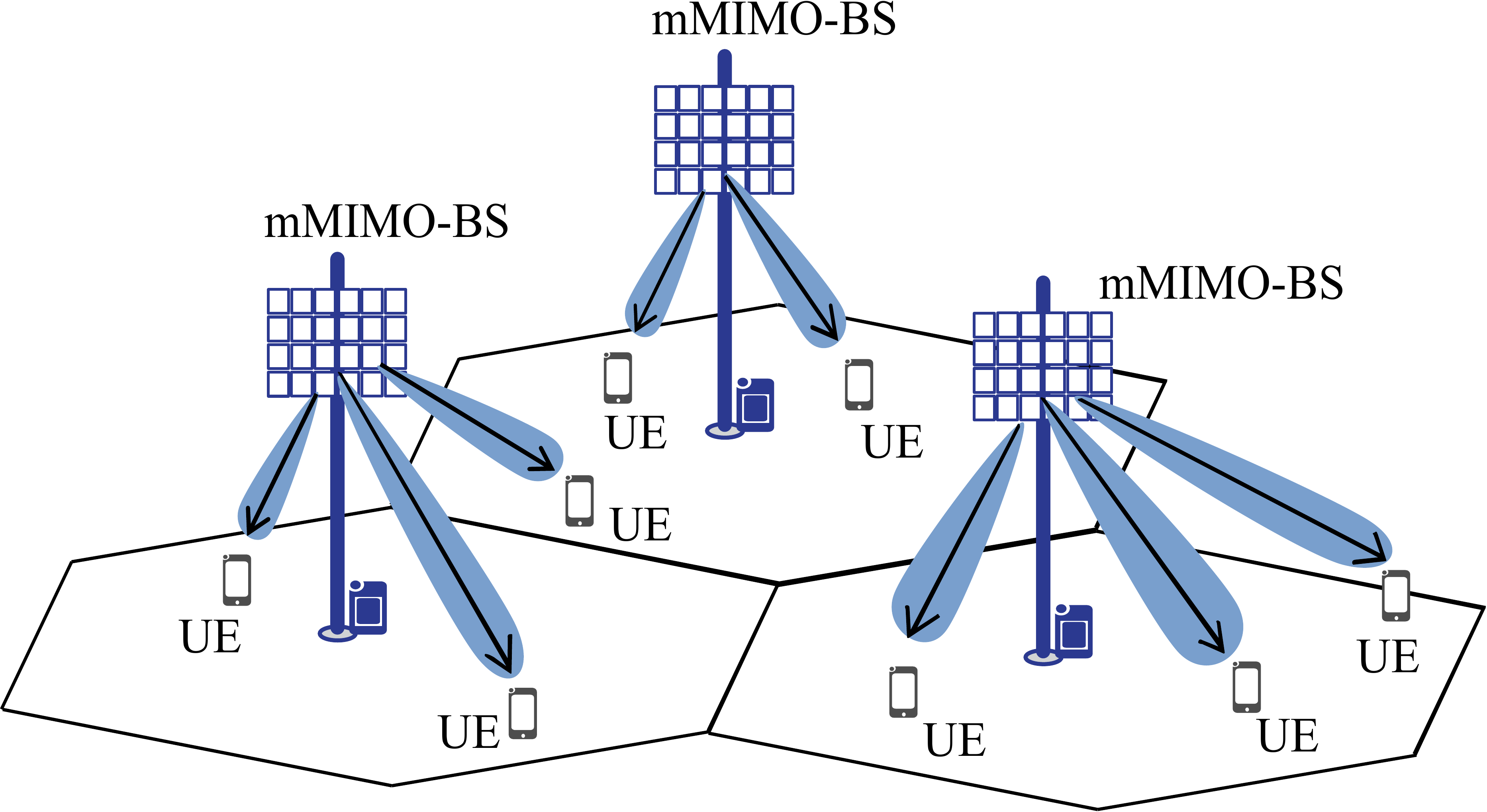}\label{fig1c}}	  
	\caption{Illustration of (a) s-BH architecture with random deployment, (b) s-BH architecture with ad-hoc deployment, and (c) DA architecture.}
	\label{fig:network_layout}
	%\vspace*{-0.3cm}
\end{figure}

We focus on the study of the DL performance for an hexagonal grid of mMIMO-BSs equipped with a large number of antennas $M$ and providing wireless backhaul links to {(a)} randomly deployed self-backhauled SCs, 
{(b)} ad-hoc deployed self-backhauled SCs, 
or {(c)} directly serving UEs,
as illustrated in Figs. \ref{fig1a}, \ref{fig1b}, and \ref{fig1c}, respectively.
Throughout the paper, we assume to use sub-6 GHz frequencies, which can provide backhaul connections with a single hop to all the SCs deployed in the coverage area of the macro cell \cite{7306534}.

For the mMIMO-BSs, we consider the hexagonal model since it represents a planned scenario of deployment where the BS locations are optimized to improve the SINR performance.
This deployment strategy is in line with the 3GPP methodology described in \cite{3gpp.36.814}, which we take as a reference throughout the paper.

For the SCs, we differentiate the modeling between the random and the ad-hoc cases. 
In the first case, we model the SC locations with a homogeneous spatial poisson-point-process (SPPP) distribution, which we refer to as SPPP, since it represents an unplanned scenario of deployment, where the SCs are distributed according to a uniform distribution.
This approach is in line with the works such as \cite{5288507,5165314}, and also with the 3GPP methodology \cite{3gpp.36.814}. 
In the second case, we consider the SC locations coupled to the UE point process, and we position the SCs at a fixed distance $d$ with respect to the UE locations.

While the random deployment of SCs is agnostic to the UE positions, the ad-hoc deployment of SCs requires knowing the position of the UEs before making the deployment. This information can be obtained through predictions, for example, considering the behavior of UEs, who are likely to return to the same place at the same time of day.

\subsection{Macro Cell, Small Cell and User Topology}\label{subs:sc-dep}
We denote by $i$, $l$, and $k$ the mMIMO-BS in the $i$-th sector, the SC, and the UE, respectively.
$\mathcal{I}$ represents the set of mMIMO-BSs deployed in the network.
$\mathcal{L}_{i}$ and $\mathcal{L}_{i'}$ represent the set of SCs deployed per sector and connected to the $i$-th and $i'$-th mMIMO-BS, respectively, 
which provides the largest reference signal received power (RSRP).\footnote{
We remark that a given SC deployed in the $i$-th sector might be connected to another mMIMO-BS $i'$ as it provides a higher RSRP level than the mMIMO-BS $i$.}
$L_{i}$ and $L_{i'}$ denote the number of SCs in the sets $\mathcal{L}_{i}$ and $\mathcal{L}_{i'}$, respectively.
Furthermore, we denote by $K_{i}$ the number of UEs in the set $\mathcal{K}_{i}$ randomly and uniformly distributed over the area covered by each sector.
We assume that each single-antenna UE is connected with the SC (in the s-BH network), or with the mMIMO-BS (in the DA network) that provides the largest RSRP \cite{3gpp.36.814}. 
Therefore each SC serves $K_{l}$ UEs in the s-BH network.

Three different network deployments are presented in the following:

\begin{enumerate}[label={(\alph*)},leftmargin=*]

\item 

	\textbf{s-BH architecture with random deployment:} Self-backhauled SCs are randomly and uniformly distributed over the mMIMO-BS geographical area,
	as shown in Fig. \ref{fig1a}. 
	This scenario is used as a baseline, 
	and follows the set of parameters specified by the 3GPP in \cite{3gpp.36.814} to evaluate the relay scenario.
	More precisely, we consider the UE and SC antenna heights fixed at 1.5 and 5 meters above the ground, respectively, and channel models for the 3GPP Case 1 Relay scenario.
\item 

	\textbf{s-BH architecture with ad-hoc deployment:} Self-backhauled SCs are positioned targeting nearby UE locations.\footnote{
	We assume the possibility to realize this specific network deployment, for example by means of drone-BSs, 
	where the drone-BSs can reposition themselves following the locations of UEs as suggested in \cite{8403635}.
	Although mentioned, the drone-BSs use-case is not the focus of this paper, and it is left for future investigation.}
	As shown in Fig. \ref{fig1b}, 
	we model this scenario by considering SCs deployed within a 2-D (two-dimensional) distance $d$ from the UEs, 
	and an angle $\psi$ measured from the straight segment that links UEs and their closest mMIMO-BS. 
	$\psi$ is chosen uniformly at random in $\left[-\pi/2,\pi/2\right]$. 
	It is worth noting that even when the 2-D distance $d=0$, 
	UEs and SCs are still separated in space because the antennas are positioned at different heights, as specified in (a).
	In addition, to limit the effect of the inter-cell interference, 
	we replace the Patch antenna at the SC with a more directive Yagi antenna, 
	pointing downwards to the ground (as shown by the green radiation cone in Fig. \ref{fig1b}), 
	and therefore only illuminating the closest UEs.
	%Details about this model can be found in Table \ref{table:parameters}. \label{ad-hoc}

\item 

	\textbf{DA architecture:} There are no self-backhauled SCs deployed, and the mMIMO-BSs are solely dedicated to directly serve the UEs, as illustrated in Fig. \ref{fig1c}.

\end{enumerate}

\subsection{Frame Structure} \label{SubSec: frame}

\begin{figure}[!t]
	\centering
	\subfloat[Self-backhaul frame]{\includegraphics[width=1\columnwidth]{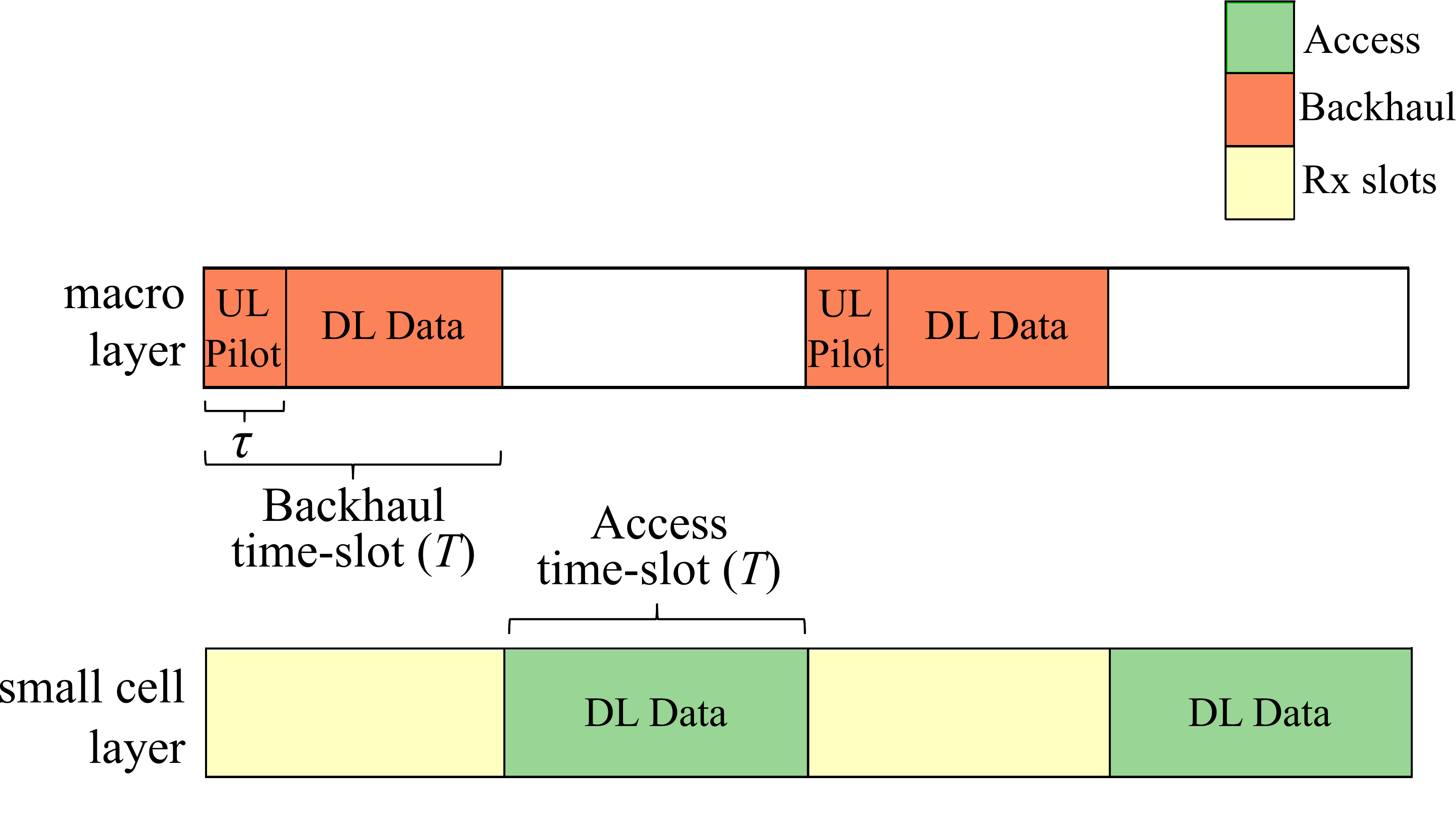}\label{fig:2a}}%
	\hfill
	\subfloat[Direct access frame]{\includegraphics[width=1\columnwidth]{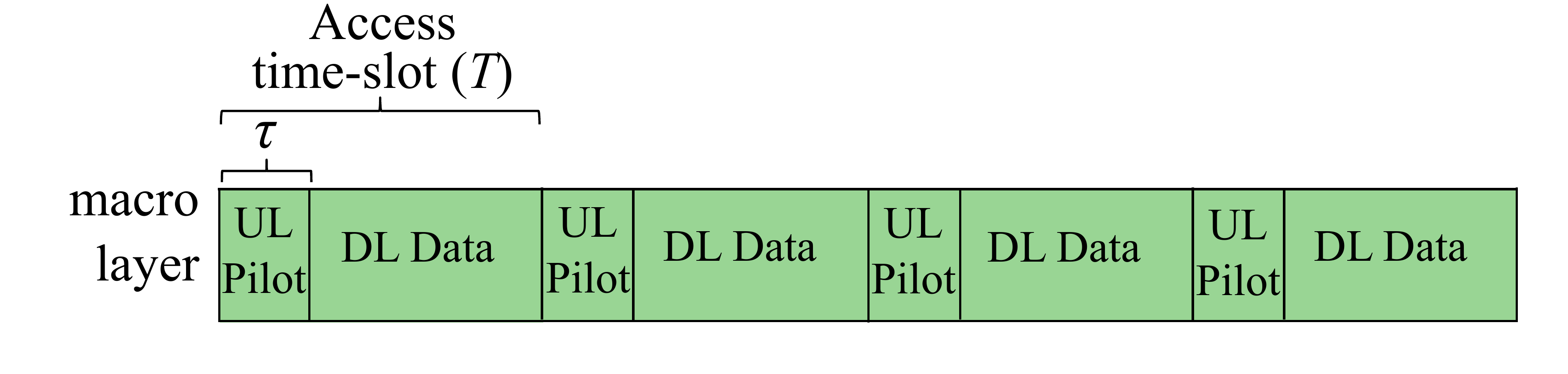}\label{fig:2b}}%
	\caption{DL frame structure for (a) mMIMO s-BH with $\alpha = 0.5$, and for (b) mMIMO DA.}
	%\vspace*{-0.2cm}
	\label{fig:frame structure}
\end{figure}

We consider a time-division duplexing (TDD) system, 
where the time-slot duration $T$ is used as a single scheduling unit in the time domain. 
As shown in Fig. \ref{fig:2a}, 
we partition the access and backhauling resources through the parameter $\alpha \in [0,1]$. 
Therefore, the fraction $\alpha$ of time-slots is allocated to the backhaul links, 
and the fraction $1-\alpha$ of time-slots is allocated to the access links. 
In the frequency domain, we divide the system bandwidth $BW$ into $Q_{t}$ RBs,
and we allocate all the RBs to the backhaul links or the access links. 
We make the following assumptions in considering the partition of backhaul and access time-slots among the SCs and UEs:

%\begin{itemize}[leftmargin=*]
\begin{itemize}
	\item
	During the backhaul time-slots,
	all the associated SCs are served by the mMIMO-BS $i$ with $M$ antennas, 
	and we use the same value of $\alpha$ for all the SCs. 
	The mMIMO-BSs precode the backhaul signals towards the single-antenna SCs, 
	which are spatially multiplexed in the same time-frequency resources.
	We consider to have $M$ DoF for a mMIMO-BS with $M$ antennas \cite{6241389},
	and we assume $M > \BBE[L_{i}]$ by dimensioning the number of antennas $M$ during the mMIMO-BS deployment planning.
	In the case that $L_i$ may exceed the number of antennas $M$, a scheduler ensures to serve only $M$ SCs randomly picked in the set $\mathcal{L}_{i}$.
	We also observed in our experiments that $\Pr[L_{i}>M]$, 
	i.e. the probability that the RV $L_{i}$ exceeds the number of DoF $M$, 
	decays rapidly for large $M$, and is equal to zero for the set of system parameters adopted in our simulations.

	\item 
	During the access time-slots, 
	the SCs schedule their connected UEs by using a Round Robin (RR) mechanism, 
	which equally distributes the available $Q_{t}$ RBs among its UEs. 
\end{itemize}

Fig. \ref{fig:2b} shows the frame structure used for the DA setup, 
where all the time-slots are allocated to the access links. 
In each time-slot, the mMIMO BSs precode the access signals,
and the UEs are spatially multiplexed reusing the entire system bandwidth.
We assume $M > \BBE[K_{i}]$ due to the system design.
When $K_{i}>M$, a scheduler serves only $M$ UEs randomly picked in the set $\mathcal{K}_{i}$.
The same observation made above for the backhaul applies with $K_{i}$ in place of $L_{i}$.

Figs. \ref{fig:2a} and \ref{fig:2b} also show the fraction $\tau$ of the time-slots dedicated for the transmission of the uplink (UL) pilot sequences, 
used for the CSI acquisition.
Details about the CSI acquisition procedure will be discussed in subsection \ref{subs:chtrain}.

\subsection{Channel Model}

We define as $\bm{\mathrm{h}}_{il}=[h_{il1},\ldots,h_{ilM}]^{\mathrm{T}} \in \mathbb{C}^{M}$ the propagation channel between the $l$-th single-antenna receiver (SC in the mMIMO s-BH network and UE in the mMIMO DA) and the $M$ antennas of the $i$-th mMIMO-BS. 
The composite channel matrix between the $i$-th mMIMO-BS and the receivers in the $i'$-th cell is represented by $\bm{\mathrm{{H}}}_{i, i'} = [\bm{\mathrm{{h}}}_{i1} \cdots \bm{\mathrm{{h}}}_{iL_{i'}}] \in \mathbb{C}^{M \times L_{i'}} $, 
where we omit the subscript $q$ indicating the q-th RB of the channel matrix for notational convenience.
Moreover, for the mMIMO sBH architecture, 
we define the single-input single-output (SISO) channel between the $l$-th SC and the $k$-th UE in the $q$-th RB as $g_{lkq} \in \mathbb{C} $. 

The channel coefficients $h_{ilm} = \sqrt{ \beta_{il}} \tilde{h}_{ilm}$ and $g_{lkq} = \sqrt{ \beta_{lk}} \tilde{g}_{lkq}$ account for both the effects of the large-scale fading and the small-scale fading components: 
%\begin{itemize}[leftmargin=*]
\begin{itemize}
	\item 
	We model the large-scale fading components $\beta_{il},\beta_{lk} \in \mathbb{R}^{+}$ according to the 3GPP Case 1 Relay scenario \cite{3gpp.36.814}.
	For a given link, the models decide whether the channel propagation conditions are LoS or NLoS, by considering a distance-dependent LoS probability function, 
	and use log-normal distributed shadowing with different values of standard deviation.
	Because of its slow-varying characteristic, 
	the large-scale fading does not change rapidly with time, 
	and it can be assumed constant over the observation time-scale of the network.
	\item 
	We model the small-scale fading components $\tilde{h}_{ilm}, \tilde{g}_{lkq} \in \mathbb{C}$, 
	which result from multi-path, 
	as Rician fast-fading, according to the 3GPP spatial channel model for MIMO simulations \cite{3gpp.25.996}, 
	assuming a $K$-factor dependent on the distance between transmitter and receiver.
\end{itemize}
	Throughout the paper we assume a block-fading channel model, 
	where the channel vectors of the access paths,
	i.e. the channel between each antenna at the mMIMO-BS and the UE, and the channel between the SC and the UE, 
	remain constant for a frequency-time block corresponding to one time-slot $T$, and one RB \cite{marzetta2016fundamentals,1237141,4277071}. 
	In line with the LTE numerology \cite{molisch2010wireless}, 
	we consider the duration of the time-slot equal to 1 ms, 
	and the bandwidth of the RB equal to 180 KHz.
	Differently, we consider that the channel vectors of the backhaul path,
	i.e. the channel between each antenna at the mMIMO-BS and the SC, 
	is constant for a period $T_{BH} \gg T$ due to the static position of the SCs. %\cite{bourdoux2015d1}
	%Moreover, we consider the Jakes correlation model to characterize the channel correlation between different antennas of the mMIMO-BS \cite{656151}.
    Moreover, we adopt a horizontal uniform linear array (ULA) for the mMIMO-BS with equally spaced antennas and Jakes correlation model between antenna pairs \cite{656151}.
    
\subsection{Massive MIMO CSI Acquisition} \label{subs:chtrain}

To calculate the DL precoder of the mMIMO-BS, 
we consider that the channel is estimated through UL pilot sequences, 
assuming UL/DL channel reciprocity \cite{6375940}. 
We also consider that the SCs or UEs associated to the same mMIMO-BS have orthogonal pilot sequences, 
and define the pilot code-book with the matrix $\bm{\mathrm{\Phi}}_{i} = [ \bm{\mathrm{\phi}}_{i1} \cdots \bm{\mathrm{\phi}}_{iL_{i}} ]^{\mathrm{T}}  \in \mathbb{C}^{L_{i} \times  B}$, 
which satisfies $ \bm{\mathrm{\Phi}}_{i} \bm{\mathrm{\Phi}}_{i}^\mathrm{H} = \bm{\mathrm{I}}_{L_{i}}$. 
Here, the $l$-th sequence is given by $\bm{\mathrm{\phi}}_{il} = [\phi_{il1},\ldots,\phi_{ilB}]^{\mathrm{T}} \in \mathbb{C}^{B}$, 
and $B$ denotes the pilot code-book length. 
Note that $L_{i} \leq B$, 
i.e. the number of SCs trained by the mMIMO-BSs in the backhaul time-slots is limited by the maximum number of orthogonal pilot sequences.
Similarly, $K_{i} \leq B$, i.e. the number of UEs trained by the mMIMO-BSs in the access time-slots is limited by the maximum number of orthogonal pilot sequences. 
In case $L_{i}>B$ or $K_{i}>B$, the scheduler assigns the pilot sequences to only $B$ SCs or UEs, randomly selected in the sets $\mathcal{L}_{i}$ and $\mathcal{K}_{i}$.\footnote{The assumption on the number of pilots holds since we focus on broadband networks and we address the conventional use case of human-type communications. Differently, the application of mMIMO to other use cases, such as Machine-Type Communication and Internet of Things, would require to revise this assumption.}

The matrix $\bm{\mathrm{Y}}_{i} \in \mathbb{C}^{M \times  B}$ of pilot sequences received at the $i$-th mMIMO-BS can be expressed as \cite{Zhu2016}
\begin{equation} \label{eq: ul}
\bm{\mathrm{Y}}_{i} = \sqrt{P_{il}^{\mathrm{ul}}} \sum\limits_{i' \in \mathcal{I}} \bm{\mathrm{{H}}}_{i,i'} \bm{\mathrm{\Phi}}_{i'} + \bm{\mathrm{{N}}}_{i},
\end{equation}
where $P_{il}^{\mathrm{ul}}$ is the power used for UL pilot transmission by the $l$-th device, located in the $i$-th sector, 
and $\bm{\mathrm{{N}}}_{i} \in \mathbb{C}^{M \times  B}$ represents the additive noise matrix, 
whose entries are modeled as independent and identically distributed complex Gaussian RVs with variance $\sigma^2$.

Let us denote by $\bm{\mathrm{{H}}}_{i} = [\bm{\mathrm{{h}}}_{i1} \cdots \bm{\mathrm{{h}}}_{iL_{i}}] \in \mathbb{C}^{M \times L_{i}} $ the channel between the $i$-th mMIMO-BS and the associated devices. 
During the UL training phase, 
the mMIMO-BS obtains an estimate of $\bm{\mathrm{{H}}}_{i}$ by correlating the received signal with a known pilot matrix $\bm{\mathrm{\Phi}}_{i}$. 
Let us define $\mathcal{P} \subseteq \mathcal{I}$ as the subset of sectors, 
whose devices share identical pilot sequences with the devices served by the $i$-th mMIMO-BS.
In line with other studies \cite{Zhu2016,6415397}, 
we adopt the least-squares (LS) channel estimation since we are considering a practical system, where the mMIMO-BS does not have perfect knowledge of the cross-cell channel statistics. The resulting LS channel estimation can be expressed as \cite{Zhu2016}
\begin{equation} \label{eq: ch est}
\bm{\mathrm{\widehat{H}}}_{i} = \frac{1}{\sqrt{P_{il}^{\mathrm{ul}}}} \bm{\mathrm{Y}}_{i}  \bm{\mathrm{\Phi}}_{i}^{\mathrm{H}} = \bm{\mathrm{{H}}}_{i} + \sum_{i' \in \mathcal{P}} \bm{\mathrm{{H}}}_{i,i'} + \frac{1}{\sqrt{P_{il}^{\mathrm{ul}}}} \bm{\mathrm{{N}}}_{i} \bm{\mathrm{\Phi}}_{i}^{\mathrm{H}}.
\end{equation}
The first, second and third terms on the right-hand side of \eqref{eq: ch est} represent the estimated channel, 
a residual pilot contamination component and the noise after the pilot sequence correlation, respectively. 
The use of the same set of orthogonal pilot sequences among different sectors leads to the well-known \emph{pilot contamination} problem, 
which can severely degrade the performance of mMIMO systems \cite{6375940, Galati1712}.

Similar to \cite{Galati1712}, 
we consider that each uplink training symbol $\tau$ contains up to 16 orthogonal pilot sequences, and we consider its duration equal to one Orthogonal Frequency Division Multiplex (OFDM) symbol period of the Long Term Evolution (LTE) standard \cite{molisch2010wireless}.
The pilot sequences are assigned to the SCs and to the UEs, in the case of mMIMO for backhaul and mMIMO for access, respectively.

In the mMIMO s-BH architecture,
due to the static position of the SCs, 
the backhaul channel vectors remain constant for a longer time than the access channel vectors, and there is no need to update the CSI every time-slot.
This allows to multiplex in time, over separate backhaul time-slots, the pilots transmitted by the SCs in different sectors.
Therefore, we assume that no pilot contamination is present in the system, and we account for a pilot overhead $\tau$ over $T$, equivalent in our system model to $7$\% of the time-slot duration.

In the mMIMO DA architecture, we use two pilot allocation schemes \cite{Galati1712}:
\begin{itemize}[leftmargin=*]
	\item 
	In \emph{pilot reuse 1 (R1)} the same set of pilot sequences is reused in all the sectors of the network, thus introducing strong pilot contamination.
	The pilot overhead to account for is $\tau$ over $T$, equivalent in our system model to $7$\% of the time-slot duration.
	
	\item 
	In \emph{pilot reuse 3 (R3)} we consider coordination between the same cell sites, avoiding using the same pilot sequence from the adjacent sectors.
	This reduces the effect of the pilot contamination at the expense of increasing the pilot overhead, which is $3 \times \tau$ over $T$, equivalent in our system model to $21$\% of the time-slot duration.
	
\end{itemize}
In Table \ref{table:pilot} we summarize the different pilot allocation schemes used in mMIMO s-BH and mMIMO DA architectures.

\begin{table}[!t]
	%\vspace{5mm} 
	\centering
	\caption{Pilot allocation schemes in mMIMO s-BH and mMIMO DA architectures.}
	%\vspace{-2mm} 
	\label{table:pilot} 
	\begin{tabulary}{\columnwidth}{|c|c|m{2.3cm}|}%{ |m{2.5cm}| m{2.3cm} | m{2.3cm} | }% double column {\columnwidth}{ |p{2.9cm}| p{5cm}  | }
		\hline
		\textbf{Architecture}		& \textbf{Backhaul links } & \textbf{Access links } \\ \hline
		mMIMO s-BH					& \multicolumn{1}{m{2.8cm}|}{pilot reuse for all the sectors (No pilot contamination)} & No pilot allocation \\ \hline
		mMIMO DA					& -- & {$\bullet$ pilot reuse 1 (R1) $\bullet$ pilot reuse 3 (R3) } \\ \hline
	\end{tabulary}
	%\vspace{-5mm} 
\end{table}

\section{Downlink SINR and User Rate}\label{Sec:3}

In this section, we present the formulation for the two-hop DL data-rate in the s-BH network, 
which comprises the formulation for the mMIMO backhaul and the SC access SINRs and data-rates.
Moreover, we include the conventional formulation for the data-rates in mMIMO DA network.

\subsection{Massive MIMO Backhaul Transmission} \label{Sec3bhTx}

The $i$-th mMIMO-BS uses the precoding matrix $\bm{\mathrm{W}}_{i}=[\bm{\mathrm{w}}_{i1} \cdots \bm{\mathrm{w}}_{iL_{i}}] \in \mathbb{C}^{M \times L_{i}}$ to serve its connected SCs during the backhaul time-slot. 
In this paper, we consider that $\bm{\mathrm{W}}_{i}$ is computed based on the zero-forcing (ZF) criterion as \cite{1261332}\footnote{As shown in \cite{massivemimobook}, among the linear methods for the mMIMO precoder, ZF and minimum mean square error (MMSE) perform better than Maximum Ratio Combining (MRC). 
However, ZF provides a better balance between performance and complexity than MMSE.}
\begin{equation} \label{eq: zf}
\bm{\mathrm{W}}_{i} = \bm{\mathrm{\widehat{H}}}_{i} \left({\bm{\mathrm{\widehat{H}}}^{\mathrm{H}}_{i}} {\bm{\mathrm{\widehat{H}}}_{i}} \right) ^{-1}{\bm{\mathrm{D}}_{i}}^{\frac{1}{2}}.
\end{equation} 
Here, the diagonal matrix $\bm{\mathrm{D}}_{i} = \operatorname{diag} \left(\rho_{i1},\rho_{i2},\ldots,\rho_{iL_{i}} \right)$ is chosen to equally distribute the total DL power $P_{i}^{\mathrm{dl}}$ among the $L_{i}$ receivers. 
In the previous expression, 
$\rho_{il}$ represents the power allocated to the $l$-th receiver located in the $i$-th sector, 
and $\Tr\{\bm{\mathrm{D}}_{i} \}=P_{i}^{\mathrm{dl}}$, 
where $\Tr\{\bm{\mathrm{D}}_{i} \}$ is the trace of matrix $\bm{\mathrm{D}}_{i}$.
%The above entails that $P_{i}^{\mathrm{dl}} = \sum_{l=1}^{L_{i}}{ \rVert \bm{\mathrm{w}}_{il} \rVert}^2$, where ${ \rVert \bm{\mathrm{w}}_{il} \rVert}$ indicates the norm of $\bm{\mathrm{w}}_{il}$. %TOREMOVE

Under the assumption that each SC has perfect CSI available, the DL SINR of the $l$-th stream transmitted by the $i$-th mMIMO-BS can be expressed as
\begin{equation} \label{eq: sinr MIMO}
\mathrm{SINR}_{il}^{\mathrm{B}} = \dfrac{ \rho_{il} | \bm{\mathrm{h}}^{\mathrm{H}}_{il} \bm{\mathrm{w}}_{il} |^2}
{ \sum\limits_{\substack{j \in \mathcal{L}_{i} \\ j\neq l}}
	{ \rho_{ij} | \bm{\mathrm{h}}^{\mathrm{H}}_{il} \bm{\mathrm{w}}_{ij} |^2} + \sum\limits_{\substack{i' \in \mathcal{I}  \\ i' \neq i}} \sum\limits_{j \in \mathcal{L}_{i'}} {\rho_{i'j} | \bm{\mathrm{h}}^{\mathrm{H}}_{i'l} \bm{\mathrm{w}}_{i'j}|^2} + \sigma^2_{n}}.
\end{equation}
The numerator of \eqref{eq: sinr MIMO} contains the power of the signal intended for the $l$-th receiver, 
and the denominator includes the intra-cell interference from the serving $i$-th mMIMO-BS, 
the inter-cell interference from other mMIMO-BSs, 
and the power of the thermal noise at the SC receiver $\sigma^2_{n}$.

The corresponding DL backhauling rate at the $l$-th SC receiver can therefore be expressed as
\begin{equation} \label{eq: rate mMIMO BH}
R_{il}^{\mathrm{B}} = \alpha \left(1 - \frac{\tau}{T} \right) BW \log_2 \left( 1 + \mathrm{SINR}_{il}^{\mathrm{B}} \right).
\end{equation}
where $\alpha$, as indicated before, represents the fraction of time-slots allocated to the backhaul links. 

\subsection{Small Cell Access Transmission}

We recall from the channel model that $g_{lkq}$ denotes the SISO channel between the $l$-th SC and the $k$-th UE corresponding to the $q$-th RB. 
The DL SINR of the $k$-th UE served by the $l$-th SC in RB $q$ can be expressed as
\begin{equation} \label{eq: SINR access}
\mathrm{SINR}_{lkq}^{\mathrm{A}} = \dfrac{ P_{l}^{\mathrm{dl}} |g_{lkq}|^2}
{ \sum\limits_{i \in \mathcal{I}} \sum\limits_{\substack{l' \in \mathcal{L}_{i}  \\ l' \neq l}} P_{l'}^{\mathrm{dl}} |g_{l'kq}|^2  + \sigma^2_{n_{2}}},
\end{equation} 
where $P_{l}^{\mathrm{dl}}$ and $P_{l'}^{\mathrm{dl}}$ are the transmit powers on the RB of the $l$-th and $l'$-th SCs, respectively, 
and $\sigma^2_{n_{2}}$ denotes the thermal noise power at the UE receiver. 
% We assume a fixed transmit power at each SC.

The corresponding DL access rate for UE $k$ served by SC $l$ can be therefore expressed as
\begin{equation} \label{eq: rate access}
R_{lk}^{\mathrm{A}} =  \left(1-\alpha\right) \frac{BW}{Q_{t}} \sum_{q=1}^{Q_{t}} x_{q}^{k} \log_2 \left( 1 + \mathrm{SINR}_{lkq}^{\mathrm{A}} \right),
\end{equation} 
where $x_{q}^{k}= 1$ if the $q$-th RB is assigned to the $k$-th UE, 
and $x_{q}^{k}= 0$ otherwise. 

The potential aggregated DL access rate provided by the $l$-th SC is $R_{l}^{\mathrm{A}} = \sum_{k=1}^{K_{l}} R_{lk}^{\mathrm{A}}$. 
However, the actual aggregated DL access rate provided by the $l$-th SC cannot be larger than the backhaul DL rate, 
which entails that $R_{l}^{\mathrm{A}} \leq R_{il}^{\mathrm{B}}$, $\forall l \in \mathcal{L}_{i}$, and $\forall i \in \mathcal{I}$. 
In this paper, we assume that the backhaul capacity is equally divided between the $K_{l}$ UEs served by the $l$-th SC.\footnote{
The assumption of equally distributed backhaul capacity among the UEs might become a drawback for the end-to-end rates, 
when UEs served by the same SC have different rate requirements in the access links, 
and in this case, the partition of the backhaul resources among the UEs could be designed according to their demands. 
This access-based partition of the backhaul resources among the UEs is not the focus of this paper, 
and its study in the context of s-BH architecture is left for future work.} 
Therefore, the resulting end-to-end access rate for the $k$-th UE can be expressed as
\begin{equation} \label{eq: rate e2e}
R_{ilk} = \min \left( \frac{R_{il}^{\mathrm{B}}}{K_{l}}, R_{lk}^{\mathrm{A}}  \right),
\end{equation}
where the minimum is computed between the backhaul rate obtained from (5), divided by the number of UEs served by the $l$-th SC, and the access rate obtained from (7).

\subsection{Massive MIMO Direct Access Transmission}

In contrast to s-BH setups, 
mMIMO systems providing DA dedicate all their time resources to multiplex DL data streams to the UEs. 
Thus, the DL access rate of the $k$-th UE served by the $i$-th mMIMO-BS can be expressed as
\begin{equation} \label{eq: rate mMIMO AC}
R_{ik}^{\mathrm{DA}} = \left(1 - \frac{\tau}{T} \right) BW \log_2 \left( 1 + \mathrm{SINR}_{ik}^{\mathrm{DA}} \right),
\end{equation}
where the estimated channel matrix $\bm{\mathrm{\widehat{H}}}_{i} = [\bm{\mathrm{\widehat{h}}}_{i1} \cdots \bm{\mathrm{\widehat{h}}}_{iK_{i}}] \in \mathbb{C}^{M \times K_{i}} $ between the $i$-th mMIMO-BS and its connected UEs is plugged into \eqref{eq: zf}, 
to subsequently derive the DL SINR in \eqref{eq: sinr MIMO}, 
assuming that each UE has perfect CSI available, 
and the access rate in \eqref{eq: rate mMIMO AC}.

\section{Analytical SIR and Average Backhaul and Access Rates}\label{Sec:4}

In the following, 
we present a tractable formulation to model the mMIMO s-BH network, 
which approximates the backhaul and access SINR and data-rate expressions.

We recall from Section \ref{Sec:2} that the mMIMO-BS locations are distributed on a hexagonal grid with density
$\lambda_{a}=3(\frac{3 \sqrt{3}}{2}R^2)^{-1}$, where $R=\frac{d_{ISD}}{\sqrt{3}}$ is the outer sector radius of the hexagonal site, and $d_{ISD}$ is the inter-site distance between two cell sites.
For the random deployment, we consider that the locations of SCs and UEs are drawn from two independent SPPPs $\Omega_b$ and $\Omega_u$ with densities $\lambda_{b}$ and $\lambda_{u}$, respectively.
For the ad-hoc deployment, we consider that the SC locations are coupled to the UE locations drawn from a SPPPs $\Omega_u$, and we position the SCs at a fixed distance $d$ with respect to the UE locations.
The mean numbers of SCs and UEs in the finite area of the sector are obtained as $\mu_{b}=\lambda_{b}/\lambda_{a}$ and $\mu_{u}=\lambda_{u}/\lambda_{a}$, respectively \cite{haenggi_2012}.
Moreover, when considering a dense deployment of SCs, there is a high probability that there are SCs unloaded, i.e. without UEs associated to them \cite{6775036}.
We capture this effect in the analysis by considering an activation probability for the SC, which can be approximated as \cite{6205422}
\begin{equation}\label{eq:activeSCprob}
p_a \approx  1 - \left(1+\frac{\lambda_{u}}{3.5 \lambda_{b}} \right)^{-3.5},
\end{equation}
where the expression $\left(1+\frac{\lambda_{u}}{3.5 \lambda_{b}} \right)^{-3.5}$ approximates the average void probability for a typical Voronoi cell \cite{6205422}.
Consequently, the locations of the active SCs can be approximated as SPPP $\tilde{\Omega}_b$ with density $\tilde{\lambda}_{b} = p_a \lambda_{b}$ derived by thinning the SC process ${\Omega}_b$ \cite{6205422}.
Differently, in the ad-hoc deployment we assume $p_a=1$, 
meaning that each SC is active and serves the nearby UE.
The mean number of active SCs in a sector is obtained as $\tilde{\mu}_{b} = \tilde{\lambda}_{b}/\lambda_{a}$.

In the following expressions, 
SIR is used to approximate the SINR, 
since in the sub-6 GHz bands, 
with a system design which assures signal coverage, 
the system operates in interference-limited conditions, 
where the power of received interference dominates the denominator of the SINR.

\subsection{Average Rates of Massive MIMO Backhaul Transmission}

We now provide an analytical model for evaluating the average data-rate of the backhaul links, 
given the SIR of a typical SC and the spatial distribution of SCs in the sector. 
Inspired by \cite{6702841}, 
we treat the SCs as UEs, 
and we extend the framework proposed in \cite{Kelif2010} to model the SIR by considering: 
\emph{i)} the effects of antenna directionality and sectorization, 
captured with the horizontal and vertical antenna patterns, 
and by modeling the co-site interference component;
\emph{ii)} the effect of the beamforming gain due to the mMIMO precoding as proposed in \cite{Tabassum}.
We make the following assumptions in our analytical backhaul model:
\begin{itemize}
	\item 
	For simplicity, the backhaul channel is statistically modeled by considering only the large-scale fading component, 
	excluding shadowing statistics.
	\item 
	We assume LoS propagation channel conditions from the serving mMIMO-BS to the SC (and from the co-site interfering mMIMO-BSs to the SC), 
	to reflect the characteristics of the backhaul link, 
	which tends to have dominant LoS conditions between SCs and the antennas of the nearest mMIMO-BSs \cite{3gpp.36.814}. 
	On the other hand, we assume all the mMIMO-BSs from the surrounding interfering sites to be in NLoS.
	\item 
	We assume that the co-channel interference from the mMIMO-BS to other served SCs can be reasonably neglected since we adopt the ZF precoder. 
	For consistency with the assumptions made in Section \ref{subs:chtrain} and in Section \ref{Sec3bhTx}, we consider that there is no pilot contamination in the backhaul links, and we consider that each SC has perfect CSI available.

\end{itemize}
%We recall from the system model that mMIMO-BSs are equipped with a large number of antennas, and we operate in a regime where the number of antennas is much larger than the number of the served SCs  ($M >> L_a$).
%In this regime, due to the mMIMO "channel hardening effect" \cite{1327795}, and for the static positions of the SCs, the backhaul channel behaves more deterministically than the access channel. 
%Therefore, in the analytical network, the backhaul channel is statistically modeled by considering only the large-scale fading component.

For the analysis that follows, 
$r$ and $\theta$ denote two independent RVs, 
which define the distance and the angle from the SC to the serving mMIMO-BS.
Note that $r$ and $\theta$ are distributed with uniform probability density functions (pdfs) $f_{R}(r)$ and $f_{\Theta}(\theta)$ in the interval $[r_{\mathrm{min}}, \frac{d_{ISD}}{2}]$ and $[-\pi/3, +\pi/3]$, respectively,
where the distance $r_{\mathrm{min}}$ denotes the minimum distance between the mMIMO-BS and the SC.\footnote{We observed that for the value of $r_{\mathrm{min}}$ that we use in our simulations, the realizations of the (modified) spatial process could be statistically considered as if they were derived from a homogeneous SPPP, since we tested the spatial randomness with the Ripley's K-function \cite{illian2008statistical}, and we verified that the theoretical values of the SPPP are contained within the region defined by the lower and upper envelopes given by the empirical results.}
By convention, 
$\theta=0$ indicates the boresight direction in the first sector, 
and such sector is denoted as $s=1$ for each hexagonal cell formed by $S$ sectors. 

The SIR of a typical SC associated to the mMIMO-BS is modeled as
\begin{equation}
{\mathrm{SIR}}^{\mathrm{B}}(r,\theta) \approx \frac{( M - \tilde{\mu}_{b} + 1)}{\tilde{\mu}_{b}} \frac{P_{i}^{\mathrm{dl}} G_{a} G_{V}(r) G_{H,1}(\theta) \beta^{\mathrm{L}}(r)}{I_{1}(r,\theta)+I_{2}(r)} ,\label{eq:sirBHapprox}
\end{equation}
where the multiplying factor ${( M - \tilde{\mu}_{b} + 1)}/{\tilde{\mu}_{b}}$ represents the beamforming gain from mMIMO precoding.\footnote{
	Only the active SCs are spatially multiplexed in the backhaul time-slots, 
	since those inactive are not required to backhaul the UEs data.}
$G_{a}$, $G_{V}(r)$ and $G_{H,s}(\theta)$ are the antenna gains of the single mMIMO-BS element, the vertical (V), and the horizontal (H) antenna patterns, respectively.
$\beta^{\mathrm{L}}(r) = A^{\mathrm{L}} r^{-\eta^{\mathrm{L}}}$ is the pathloss between mMIMO-BS and SC, 
where $A^{\mathrm{L}}$ and $\eta^{\mathrm{L}}$ indicate the frequency dependent pathloss factor and the pathloss exponent for the backhaul link in LoS condition, respectively, 
and $I_{1}(r,\theta)$ and $I_{2}(r)$ are the co-site and inter-site interference components, respectively.
The vertical antenna pattern is defined as \cite{3gpp.36.814}
\begin{align}
G_{V}(r)|_{\mathrm{dBi}}=-\min \Bigg( {12 \Bigg( \frac{\mathrm{atan}\Big( \frac{\delta_{a}}{\sqrt{ r^2 - \delta_{a}^2}} \Big) - \zeta_{\mathrm{tilt}}}{\zeta_{HP}}
	 \Bigg)^2}, F_v \Bigg),
\end{align}
where $\delta_{a}$ is the difference in antenna heights between the mMIMO-BS and the SC, 
$\zeta_{\mathrm{tilt}}$ is the mechanical downtilt, 
$\zeta_{HP}$ is the half-power vertical beamwidth, 
and $F_v$ is the vertical front-back ratio.
Similarly, the horizontal antenna pattern is defined as \cite{3gpp.36.814} 
\begin{equation}
G_{H,s}(\theta)|_{\mathrm{dBi}}=-\min\left(12 \left(\frac{\theta - (s-1)2\pi/3 }{\theta_{HP}}\right)^2, F_h\right) , 
\end{equation}
where $\theta_{HP}$ is the half-power horizontal beamwidth, 
and $F_h$ is the horizontal front-back ratio.

In \eqref{eq:sirBHapprox}, 
$I_{1}(r,\theta)$ is represented as
\begin{equation}
I_{1}(r,\theta) = P_{i'}^{\mathrm{dl}} G_{a} G_{V}(r) \sum_{s=2}^{S} G_{H,s}(\theta) \beta^{\mathrm{L}}(r),
\end{equation}
and $I_{2}(r)$ is approximated as \cite{Kelif2010}
\begin{multline}
I_{2}(r) \approx \frac{2 \pi \lambda_{a} P_{i'}^{\mathrm{dl}} G_{a} G_{V}(2 R_c - r) \overline{G_{H}} A^{\mathrm{NL}}}{\eta^{\mathrm{NL}}-2} \\
\times \left( \left( 2 R_c - r \right)^{2-\eta^{\mathrm{NL}}} - \left( R_b - r \right)^{2-\eta^{\mathrm{NL}}} \right),
\end{multline}
where $R_b=\frac{3}{2}d_{ISD}$ denotes the network boundary, 
$R_c=\frac{d_{ISD}}{2}$ is the inner sector radius, 
$\overline{G_{H}} = \int_{0}^{2 \pi} \sum_{s=1}^{S} G_{H,s}(\theta) d\theta $ is the average horizontal antenna gain with respect to $\theta$, 
and $\eta^{\mathrm{NL}}$ and $A^{\mathrm{NL}}$ are the pathloss exponent and the frequency dependent pathloss factor for the backhaul link in NLoS condition, respectively.
Eq. (15) approximates the integration area used to calculate the inter-cell interference from other macro-BSs assuming a ring formed by two concentric discs centered at the origin and with radii of $2 R_c$ and $R_b$. 
We refer to \cite{Kelif2010} for the validation of this model.

Finally, the average SC data-rate for backhaul transmission can be expressed as
\begin{multline}
\overline{R^{\mathrm{B}}} = \alpha \, BW \\
\times \int_{-\frac{\pi}{3}}^{\frac{\pi}{3}}
\int_{r_{\mathrm{min}}}^{R_c} \log_2 \left( 1 + {\mathrm{SIR}}^{\mathrm{B}}(r,\theta) \right) f_{R}(r) f_{\Theta}(\theta)\,dr\, d\theta . \label{eq:avgBackhaul}
\end{multline}
Therefore, the results for the average SC data-rate can be computed by numerical integration of \eqref{eq:avgBackhaul}.

\subsection{Average Rates of Small Cell Access Transmission}

Inspired by the stochastic geometry analysis presented in \cite{5621983}, 
we now provide an analytical model for evaluating the access SIR of a typical UE at the origin, and its access average DL data-rate.
Similarly to \cite{7335646,7248759}, 
we consider the impact of the LoS and NLoS pathloss characteristics to model the SIR.
We use the same LoS probability function as in \cite{7248759}, however, we consider for the inter-cell interference computation the density of the active SCs, i.e. those with UEs associated.
%Differently from \cite{7335646}, 
%we consider that the serving SC always has a LoS path to the UE, 
%due to the proximity of SC to UE.
%We will show by simulations in Fig. \ref{fig:losProb_sBHonly} that this \textcolor{red}{assumption} is realistic in the considered range of SCs densities.

We make the following assumptions in our analytical access model:
\begin{itemize}
	\item 
	We assume that each UE connects to the nearest SC.\footnote{This assumption holds in the considered range of SC densities, for which the probability that the closest SC is in LoS is very high, as shown by the simulation results in Fig. 3, due to the proximity of SC to UE.}
	For the random deployment, since $\Omega_b$ and $\Omega_u$ are two independent SPPPs, we assume that $x$, i.e. the distance between UE and the serving SC is a RV Rayleigh distributed with pdf $f_X(x) = 2 \pi \lambda_{b} x \, {\exp(- \lambda_{b}  \pi x^2 )}/{\exp(- \lambda_{b} \pi \delta_{b}^2 )}$, 
	where $\delta_{b}$ denotes the difference between the SC and UE heights \cite{5621983}.
	Differently, for the ad-hoc deployment, we assume $x=\sqrt{d^2 + \delta_{b}^2}$.
	\item 
	The propagation channels are represented with a combination of distance-dependent pathloss and multi-path fading, 
	distributed as Rayleigh with an exponential power distribution $|g|^2 \sim \exp(1)$.\footnote{Shadowing statistics are neglected in the analytical model, although a more comprehensive framework can incorporate this effect in the distribution of the UE distances.}
	\item 
	We adopt a probabilistic LoS channel model for the inter-cell interference, 
	with a LoS probability expressed as \cite{7248759}
	\begin{equation}
	{\Pr}^{\mathrm{L}}(x)=\exp( - \left( {x}/{D} \right)^{2}),
	\end{equation}
	where the parameter $D$ is set to approximate the LoS probability of the SC-UE 3GPP model \cite{7248759}.
\end{itemize}

The SIR of a typical UE at the origin associated to the SC is modeled as
\begin{equation}\label{eq:sirACapprox}
{\mathrm{SIR}^{\mathrm{A}}}(x) \approx \frac{P_{l}^{\mathrm{dl}} G_{b} |g|^2 \beta^{\mathrm{L}}(x)}{I_{{\mathrm {agg}}}},
\end{equation}
where $G_{b}$ is the SC antenna gain, 
$|g|^2$ is the multi-path channel gain, 
and $I_{{\mathrm {agg}}}$ is the aggregated inter-cell interference. 

We now use \eqref{eq:sirACapprox} to provide an expression for the rate coverage probability, 
which defines the probability that the UE rate is higher than a minimum target $R_{th}$. 
This probability can be expressed as $\Pr \left[{\mathrm{SIR}^{\mathrm{A}}}(x)>\gamma_a \right]$, 
where  $\gamma_a = 2^{R_{th}\mu_{l}}-1$ depends on $R_{th}$, 
and we approximate the number of UEs associated with the SC serving the typical UE at the origin with the corresponding
mean $\mu_{l}$, given by $1 + 1.28\lambda_{u}/\lambda_{b}$ \cite{6497002}.
We later show in Fig. \ref{fig:avgNumUEperSC} that the mean of the distribution of the number of UEs for a typical SC, given by $\lambda_{u}/\tilde{\lambda}_{b}$ \cite{8515110}, matches the numerical results.

The expressions used to evaluate the rate coverage probability are included in Appendix \ref{appendix}.
Thus, the average UE data-rate for access transmission can be expressed as
\begin{multline}
\overline{R^{\mathrm{A}}} = \left( 1 - \alpha \right) BW \\
\times \int_{0}^{+\infty}\!\!
\int_{\delta_{b}}^{+\infty} \Pr \left[{\mathrm{SIR}^{\mathrm{A}}}(x)>\gamma_a \right]  f_X(x)\,dx\,d\gamma_a.\label{eq:avgAccess}
\end{multline}
Therefore, the results for the average UE data-rate can be computed by numerical integration of \eqref{eq:avgAccess}.

In the next section, 
we will use this model to complement the insights given by the 3GPP-based system-level simulations.

\section{Simulations and Numerical Results}\label{Sec:5}

\begin{table}[!t]
	%\vspace{5mm} 
	\centering
	\caption{3GPP-based system-level simulation parameters}
	%\vspace{-2mm} 
	\label{table:parameters} 
	\begin{tabulary}{\columnwidth}{ |p{2.9cm}| p{5cm}  | }% double column {\columnwidth}{ |p{2.9cm}| p{5cm}  | }
		\hline
		\textbf{mMIMO-BSs}		                    & \textbf{Description} \\ \hline
		Cellular layout				                & Wrap around hexagonal, 19 sites, 3~sectors/site \\ \hline
		Deployment				                    & Inter-site distance: $500$~m, height: $32$~m \\ \hline
		Antenna array 				                & Uniform Linear Array (ULA) with element spacing $0.5\lambda$ and Jakes correlation model \cite{656151}, Number of antennas per array: 64 \\ \hline
		Antenna pattern 			                &  $70^{\circ}$ H x $10^{\circ}$ V beamwidths, 14 $\mathrm{dBi}$ max., downtilt: $15^{\circ}$ \\ \hline
		Precoder					                & Zero-forcing \\ \hline
		Tx power/Noise figure		                & 46 dBm, 5 dBm  \\ \hline 
		\textbf{Self-BH SCs} 		                & \textbf{Description} \\ \hline
		Deployment  				                & Random: $\{4,8,16\}$ SCs/sector on average, Ad-hoc: 16 SCs/sector on average, height: $5$~m \\ \hline
		Backhaul antenna pattern 	                & 5 $\mathrm{dBi}$ antenna gain,  Omni \\ \hline
		Access antenna pattern  -- Patch	        &  $80^{\circ}$ H x $80^{\circ}$ V beamwidths, 5 $\mathrm{dBi}$ max., downtilt: $90^{\circ}$ \\ \hline
		Access antenna pattern -- Yagi	            & $58^{\circ}$ H x $47^{\circ}$ V beamwidths, 10 $\mathrm{dBi}$ max., downtilt: $90^{\circ}$ \\ \hline
		Tx power/Noise figure		                & 30 dBm, 5 dB \\ \hline
		\textbf{UEs}				                & \textbf{Description} \\ \hline
		Deployment 				                    & Random, 16 UEs/sector on average, all served, height: $1.5$~m \\ \hline
		Tx power/Noise figure     	                & 23 dBm, 9 dB \\ \hline
		\textbf{Channel} 			                & \textbf{Description} \\ \hline
		Scenario 				                    & Outdoor SCs, outdoor UEs \\ \hline
		Bandwidth/Time-slot 		                & 10 MHz at 2 GHz, $Q_{t} = 50$ RBs, $T=1$ msec. \\ \hline
		LoS probability, pathloss and shadowing	    & 
		\vspace{-\topsep}%
		\begin{itemize}[leftmargin=*]
			\item mMIMO-BS to UE (based on 3GPP macro to UE models as per \cite{3gpp.36.814})
			\item mMIMO-BS to SC (based on 3GPP macro to relay models as per \cite{3gpp.36.814})
			\item SC to UE (based on 3GPP relay to UE models as per \cite{3gpp.36.814}) \vspace*{-\baselineskip}
		\end{itemize}%
		\\ \hline
		Fast fading  				                & Rician, distance-dependent K factor \\ \hline
		Thermal noise 				                & -174 dBm/Hz power spectral density \\ \hline
	\end{tabulary}
	%\vspace{-5mm} 
\end{table}

In this section, we evaluate the performance of mMIMO s-BH and DA networks using 3GPP-based system-level simulations and mathematical analysis.

In the mMIMO s-BH network, 
the different characteristics of the backhaul and access radio links are modeled considering the methodology described in \cite{3gpp.36.814} for the 3GPP Case 1 Relay scenario. 
As described in \cite[Tab. A.2.1.1.2-3]{3gpp.36.814}, 
we adopt the LoS and NLoS pathloss exponents $\eta^{\rm{L}}=2.35$ and $\eta^{\rm{NL}}=3.63$ for the backhaul links, 
and we consider $\eta^{\rm{L}}=2.09$ and $\eta^{\rm{NL}}=3.75$ for the access links. 
For each link, 
we use the corresponding LoS probability function proposed in \cite[Tab. A.2.1.1.2-3]{3gpp.36.814}.
To simulate the backhaul links, 
we account for the SC site planning correction factor, 
which affects the pathloss and the LoS probability as indicated in \cite[Tab. A.2.1.1.4-2]{3gpp.36.814}.
To simulate the access links, 
we assume cross-correlated shadowing, 
with correlation coefficient $\rho=0.5$ at the UE location with respect to the different SCs \cite[Tab. A.2.1.1.2-3]{3gpp.36.814}.
In the simulations, 
we consider a Rician fading model, 
and we characterize the Rician $K$-factor with the model: 
$K [\mathrm{dB}] = 13-0.03r$  in dB,
where $r$ is the distance between transmitter and receiver in meters \cite{3gpp.25.996}.

In the 3GPP-based system-level simulations, 
the channel gains (composed by pathloss, shadowing and multi-path fading) are generated for all useful and interfering radio links between each SC and the UEs, 
as well as between each mMIMO-BS and all SCs. 
We collect statistics for different network realizations, 
each with independent deployments of UEs and SCs. 
Subsequently, we measure the performance in terms of the cumulative distribution functions (CDFs) 
of the backhaul SC rates in \eqref{eq: rate mMIMO BH},
of the access UE rates in \eqref{eq: rate access}, 
and of the end-to-end UE rates in \eqref {eq: rate e2e}. 

To compare mMIMO s-BH against mMIMO DA architectures, 
we also simulate the links between mMIMO-BSs and UEs, 
and compute the resulting rates in \eqref{eq: rate mMIMO AC}.
In the mMIMO DA network, 
we adopt the LoS probability function and the corresponding exponents $\eta^{\rm{L}}=2.42$ and $\eta^{\rm{NL}}=4.28$, 
as indicated in \cite[Tab. A.2.1.1.2-3]{3gpp.36.814}.

Table \ref{table:parameters} contains the relevant parameters used to conduct the simulation campaign.

\subsection{Massive MIMO s-BH: Random vs. Ad-hoc Small Cell Deployments} \label{sec:adhocvsrandom}

\begin{figure}[!t]
	\centering
	\includegraphics[width=1\columnwidth]{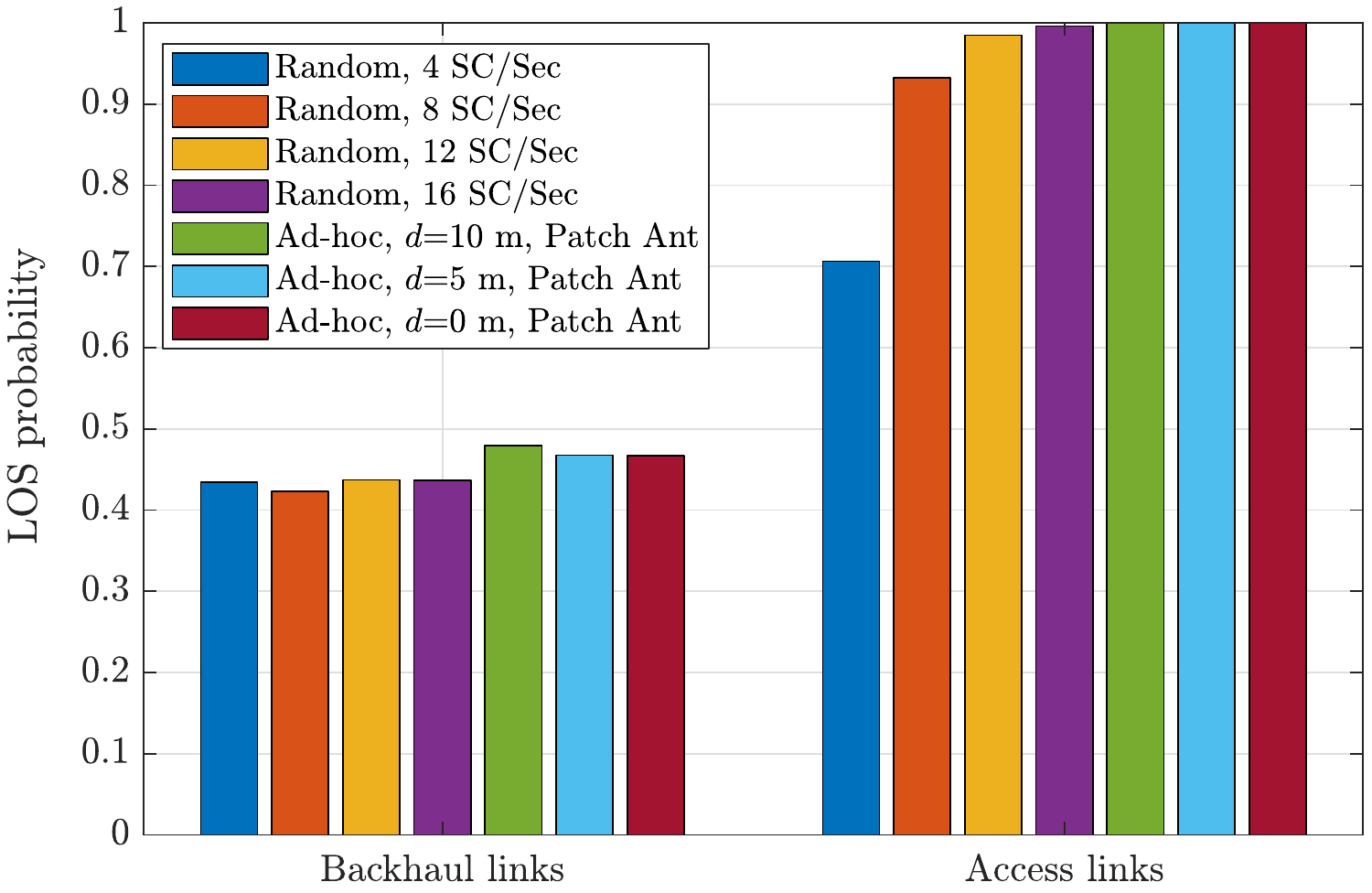}%
	\caption{LoS probability for SC backhaul links and UE access links in s-BH networks with Random and Ad-hoc deployments.}%
	\label{fig:losProb_sBHonly}%
\end{figure}

%\begin{figure*}[!t]
%	\centering
%	\subfloat[CDF of SC rates for backhaul links]{{\includegraphics[width=0.90\columnwidth]{figures/cdfBackhaulSCrate} }\label{fig:cdfBackhaulSC}}%
%	\subfloat[CDF of UE rates for access links]{{\includegraphics[width=0.90\columnwidth]{figures/cdfUEaccessRate} }\label{fig:cdfAccessRate}}\\
%	\caption{CDF of SC and UE rates for backhaul and access links.}%
%	\subfloat[Average number of SCs served per mMIMO-BS]{{\includegraphics[width=0.90\columnwidth]{figures/avgNumSCperBS} }\label{fig:avgNumSCperBS}}%
%	\subfloat[Average number of UEs served per SC]{{\includegraphics[width=0.90\columnwidth]{figures/avgNumUEperSC} }\label{fig:avgNumUEperSC}}%
%	\caption{Average number of SCs and UEs served in the backhaul and access time-slots, respectively. In \ref{fig:avgNumSCperBS} and \ref{fig:avgNumUEperSC}, the analytical results from \eqref{eq:activeSCprob} are shown to verify the approximation.}%
%\end{figure*}

\begin{figure*}[!t]
	\centering
	\subfloat[CDF of SC rates for backhaul links]{{\includegraphics[width=0.95\columnwidth]{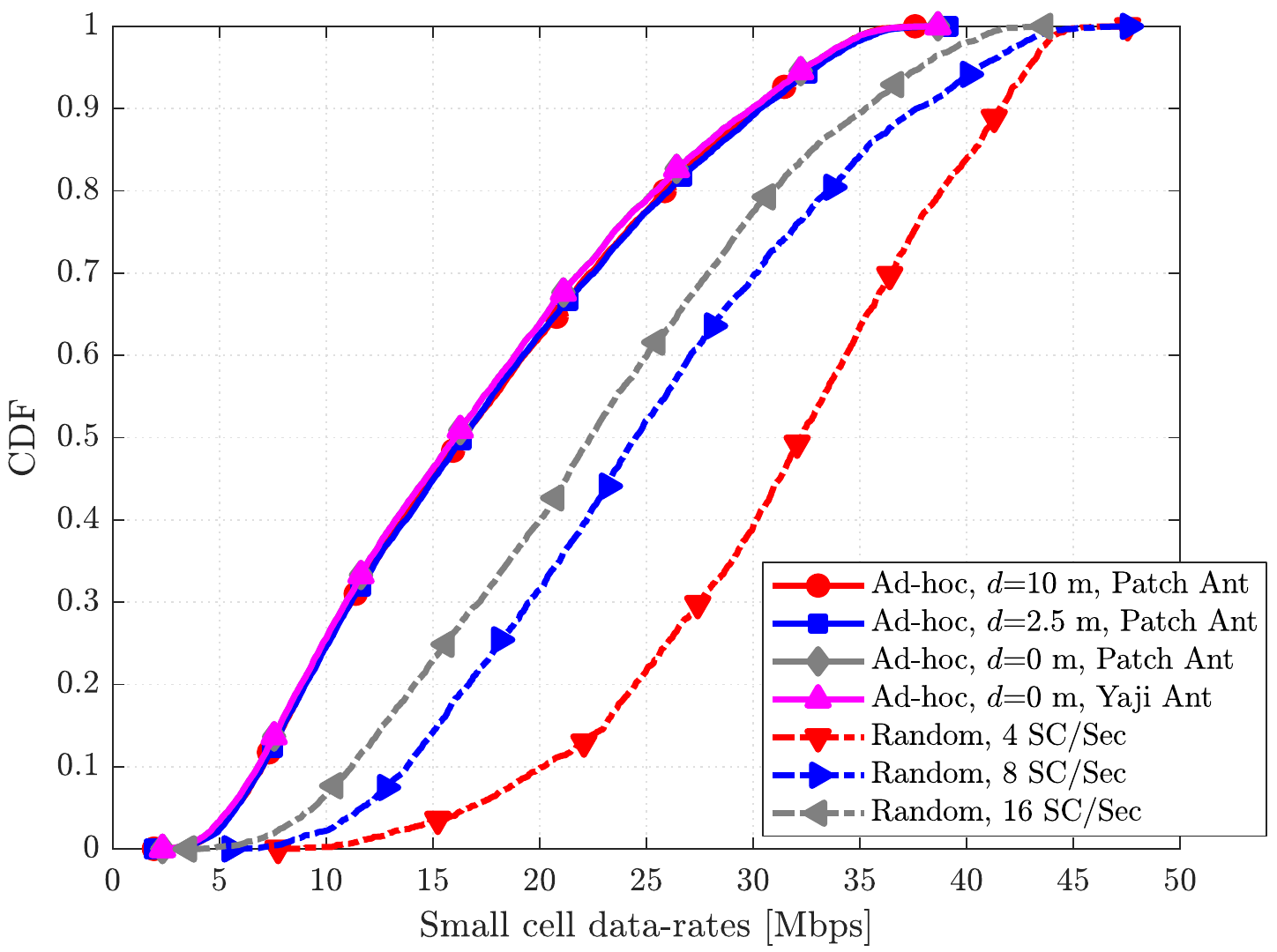} }\label{fig:cdfBackhaulSC}}%
	\hspace{0.10in}
	\subfloat[Average number of active SCs served per mMIMO-BS]{{\includegraphics[width=0.95\columnwidth]{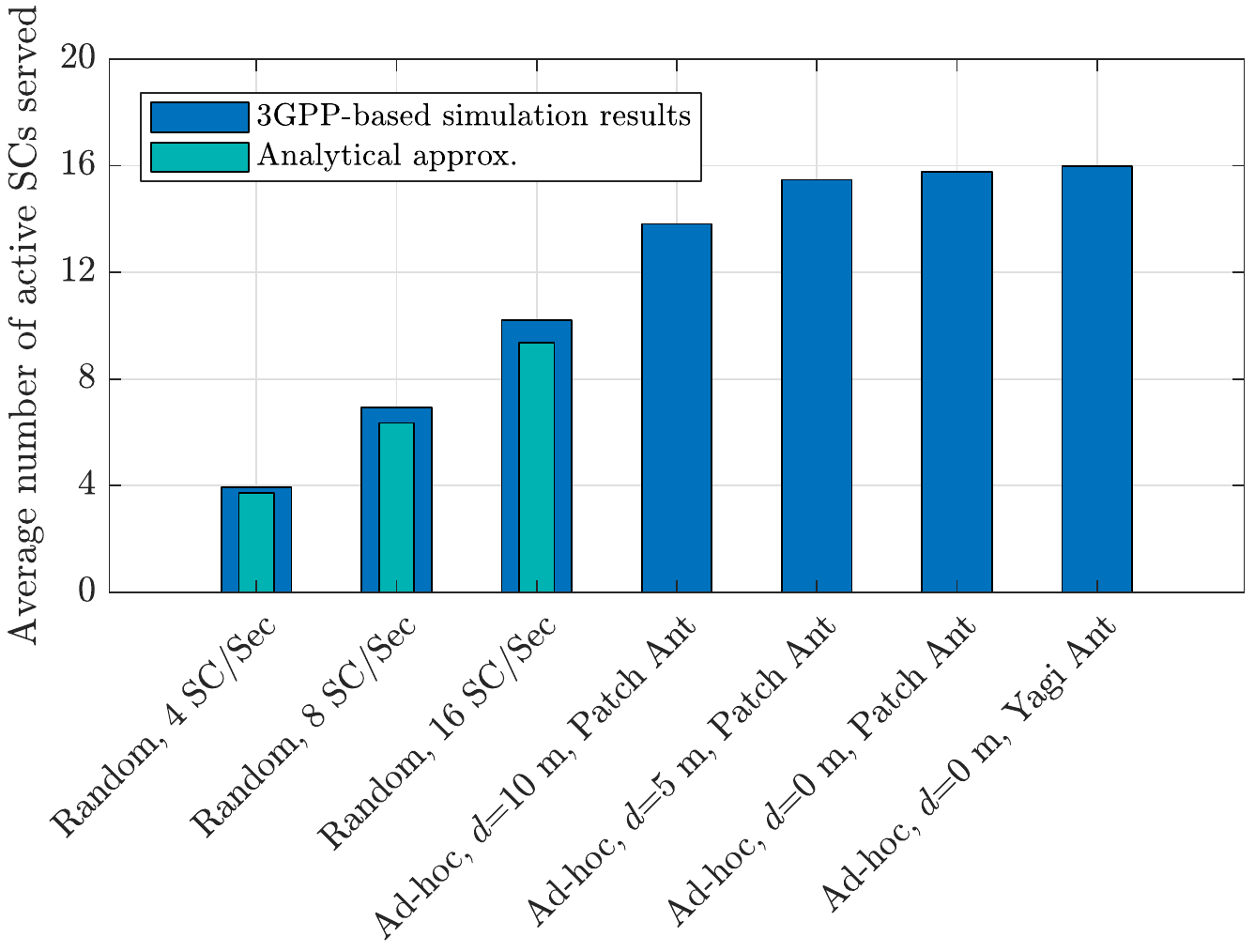} }\label{fig:avgNumSCperBS}}%
	\caption{(a) CDF of SC rates for backhaul links, and (b) average number of active SCs served in the backhaul time-slots. (b) also shows the analytical results for the SCs activation probability in \eqref{eq:activeSCprob}.}%
	\subfloat[CDF of UE rates for access links]{{\includegraphics[width=0.95\columnwidth]{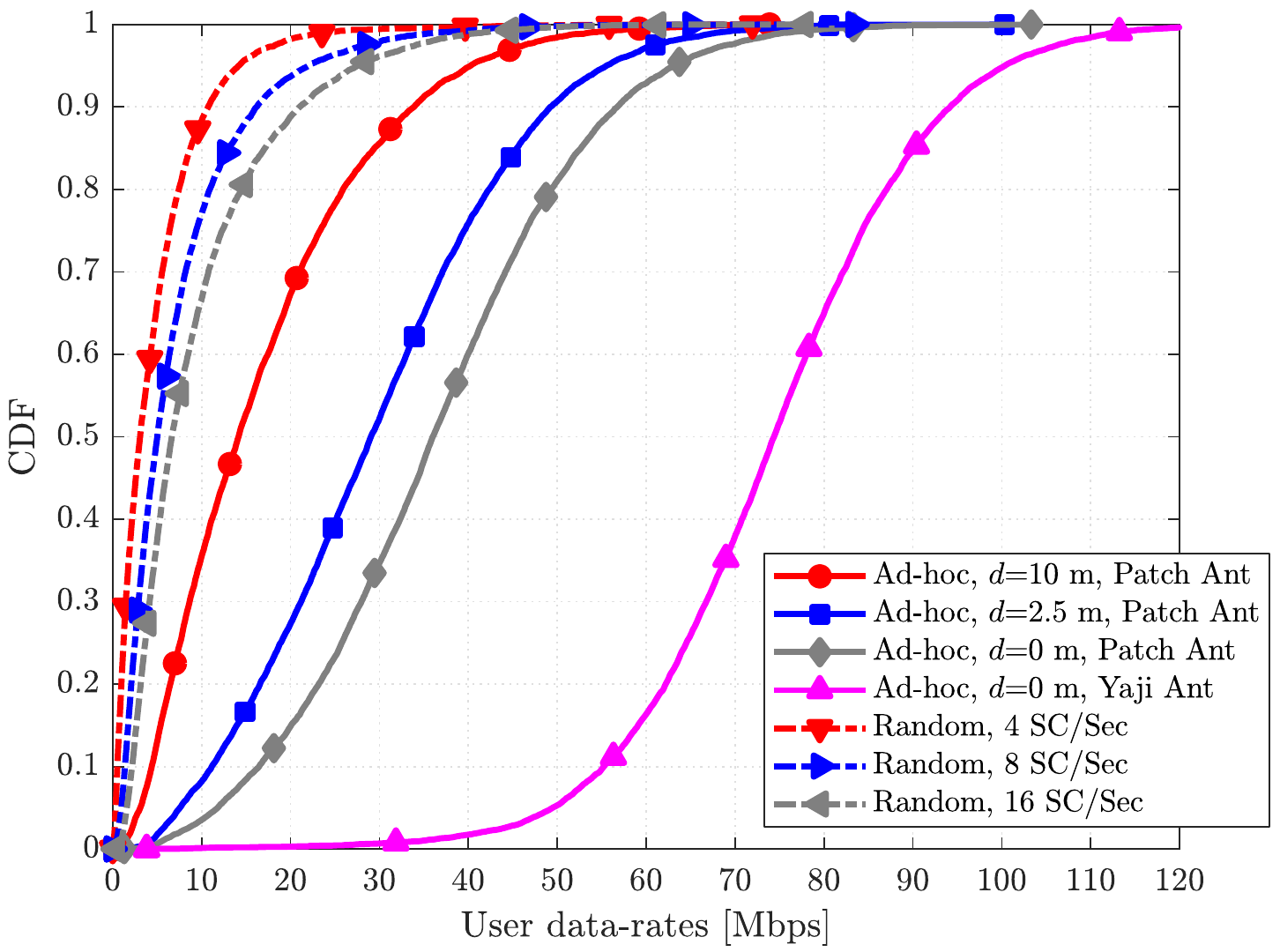} }\label{fig:cdfAccessRate}}%
	\hspace{0.10in}
	\subfloat[Average number of UEs served per active SC]{{\includegraphics[width=0.95\columnwidth]{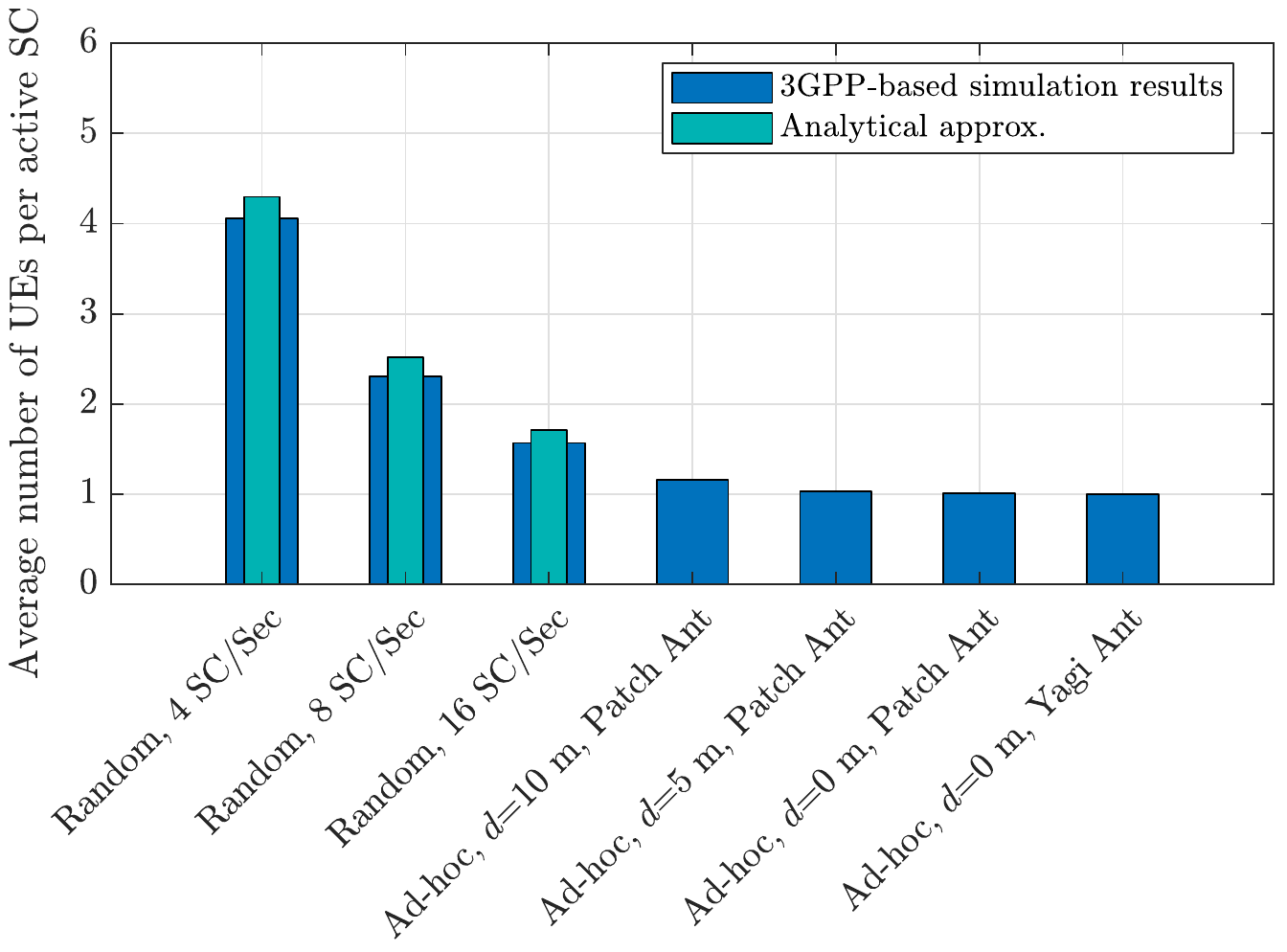} }\label{fig:avgNumUEperSC}}%
	\caption{(a) CDF of UE rates for access links, and (b) average number of UEs served in the access time-slots. (b) also shows the analytical results of the mean number of UEs served by an active SC, given by $\lambda_{u}/\tilde{\lambda}_{b}$.}%
\end{figure*}

\begin{figure}[!t]
	\centering
	%\hspace{0.10in}
	\includegraphics[width=0.90\columnwidth]{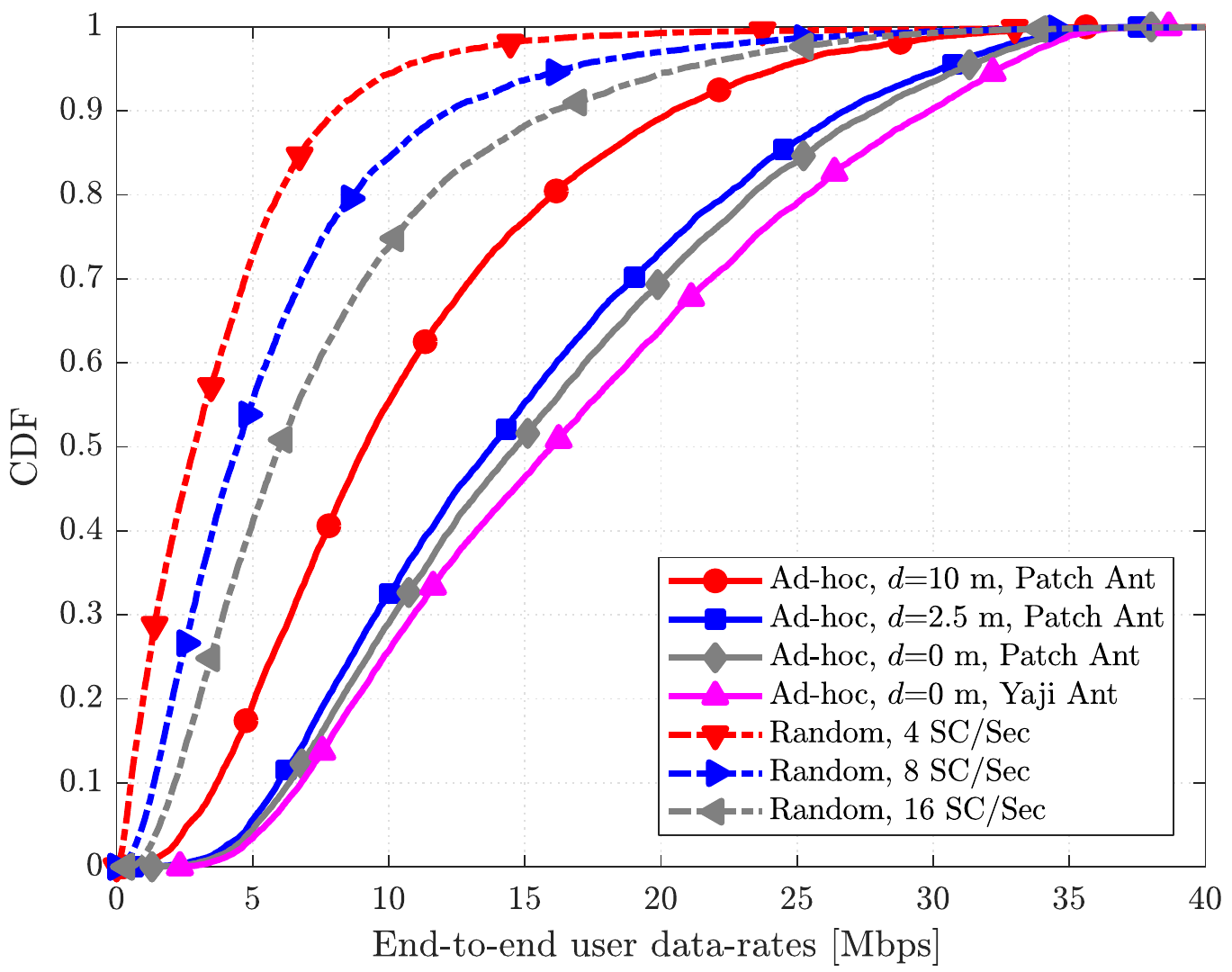}%
	\caption{CDF of end-to-end UE rates in: (i) ad-hoc deployment of 16 SCs per sector with variable UE-to-SC distance $d$, and (ii) random deployment of SCs.}%
	\label{fig:cdfE2Erate}%
\end{figure}

In this subsection, 
we analyze the 3GPP-based simulation results for the two SC topologies described in Sec. \ref{subs:sc-dep}, 
namely the ad-hoc and random SC deployments. 
In both cases, $\mu_{u}=16$ UEs are deployed per sector on average and scheduled in access time-slots by their serving SCs. 
We evaluate the impact of densification by considering $\mu_{b}=\{4,8,16\}$ SCs per sector on average for the case of random SC deployments. 
In the ad-hoc deployment,
 we consider $\mu_{b}=16$ SCs per sector on average, 
 and different values of the 2-D distance $d$ from the UE to the SC.
The resource partition $\alpha$ is set to $0.5$,
to distribute between backhaul and access the available resources equally.

As a first step, 
we compare the LoS probability of the backhaul and access links.
The group of results in the left part of Fig. \ref{fig:losProb_sBHonly} shows the probability of a given SC to be in LoS with respect to the server mMIMO-BS with different densities of SCs and deployments. 
As expected, the percentages of backhaul links in LoS are almost the same in both the random and ad hoc deployments, 
since in the first case the SCs are randomly distributed with respect to mMIMO-BSs, 
while in the last approach the SCs are positioned in the vicinity of the UEs, 
which are randomly distributed with respect to mMIMO-BSs.
Moreover, we can also see that the LoS probability of a UE in the s-BH architecture increases as the density of SCs increase, 
reaching $100\%$ probability of LoS channel condition in the ad-hoc deployment, 
as shown by the results in the right part in Fig. \ref{fig:losProb_sBHonly}.
Overall, %the results %in Fig. \ref{fig:losProb_sBHonly} 
the backhaul link mainly limits the joint backhaul-access probability of LoS-LoS conditions.

As a second step, 
Figs. \ref{fig:cdfBackhaulSC}, \ref{fig:cdfAccessRate} and \ref{fig:cdfE2Erate} analyze the data-rate performance of the backhaul and access transmission by first considering the two links separately, 
and then their combined effect on the end-to-end UE rate. 
The following considerations can be made:
%\begin{itemize}[leftmargin=*]
\begin{itemize}
	\item 
	\textit{Backhaul link performance}: Fig. \ref{fig:cdfBackhaulSC} illustrates the CDFs of the backhaul data-rates.
	These results show how increasing the number of SCs randomly deployed, and especially with the ad-hoc deployment, the backhaul data-rate received by each SC decreases. 
	This is due to the reduction of the multiplying factor ${\left[( M - \tilde{\mu}_{b} + 1)P_{i}^{\mathrm{dl}}\right]}/{\tilde{\mu}_{b}}$ in \eqref{eq:sirBHapprox}, 
	and the split of the transmit power among the active backhaul streams.
	It is worth to note that only the SCs with associated UEs are active (i.e. transmitting to the UEs), 
	and are served via multiple backhaul links. 
	Thus, looking at Fig. \ref{fig:avgNumSCperBS}, 
	which show the average number of SCs served by the mMIMO-BS when applying the random and ad-hoc SCs deployment strategies, 
	we can better explain the results presented in Fig \ref{fig:cdfBackhaulSC}. 
	In fact, while with the ad-hoc deployment almost all the 16 SCs are always active, 
	with 16 randomly deployed SCs only 10 of them are active in average, 
	as a result of the UEs association procedure. 
	%This effect justifies the trend of the backhaul data-rate, 
	%indicating how the poor performance of the backhaul in the ad-hoc deployment may constitute the main bottleneck in the two-hop communication.
	%for its improved access link capacity, boosted by the proximity and LoS conditions of the SC-to-UEs links. 
	As depicted in Fig. \ref{fig:avgNumSCperBS}, 
	the analytical approximation in \eqref{eq:activeSCprob} of the SC activation probability matches the numerical results obtained by simulations.
	
	\item 
	\textit{Access link performance}: 
	Fig. \ref{fig:cdfAccessRate} shows the results for the access data-rate. 
	As a general conclusion we can see that 
	adding more randomly deployed SCs in the sector does not introduce a significant gain, 
	while opportunistically deploying one SC closer to each UE is quite beneficial.
	In the following, we discuss the details of the different factors playing a key role in these results.
	 
	On the one hand, 
	when densifying the network, 
	the carrier signal benefits from having SCs that are more likely in close vicinity to the served UE, 
	even if a random SC deployment does not always guarantee this vicinity. 
	Also, each SC has to serve a progressively reduced number of UEs in the access links (as indicated in Fig. \ref{fig:avgNumUEperSC}), 
	and accordingly in the backhaul links, 
	which means having more RBs available to allocate to each UE over different links.
	
	On the other hand, 
	adding more SCs increases the probability of having a larger number of LoS interferers at the UE side. 
	In the random deployment, 
	the power of the interference links grows faster than the carrier signal power due to NLoS to LoS transition of the interference links \cite{7335646}. 
	In the ad-hoc case, 
	the same interference effect takes place. 
	However, by decreasing the distance $d$ from UE to SC in such a way as to ensure that the position of the SC is always close to the served UE, 
	the power of the carrier signal increases faster than the interference power,
	and thus the hit in the SINR is not as significant. 
	As a result, only a very dense deployment of random SCs could provide the same data-rate as in the case of the ad-hoc deployment. 
	In subsection \ref{subs:asymptoticRegime}, we will discuss the asymptotic behavior when increasing the density of SCs, 
	and quantify the number of required randomly deployed SCs to achieve this condition.

	From Fig. \ref{fig:cdfAccessRate}, 
	we can observe how equipping the SC with a more directive antenna (i.e. Yagi) with respect to the Patch antenna 
	further boosts the access link capacity to achieve 75 Mbps per UE at the median value.
	Two complementary effects cause this performance enhancement: 
	\emph{i)} the signal improvements provided by the higher antenna gain of the directive Yagi, 
	and \emph{ii)} the reduced interference created towards neighboring UEs served by other SCs. 
	%This is only applicable to the ad-hoc deployment, 
	%where the UEs are in close proximity to their serving SC, 
	%and fall under its antenna coverage pattern.

	\item 
	\textit{End-to-end overall performance}\label{end-to-end}: 
	Fig. \ref{fig:cdfE2Erate} shows the end-to-end results given by the combination of the two-hop, backhaul and access, performance previously depicted 
	in Figs. \ref{fig:cdfBackhaulSC} and \ref{fig:cdfAccessRate}. 
	Overall, the end-to-end data-rates of the random deployment are more limited by the access links than by the backhaul links, 
	as shown by comparing the results in Figs. \ref{fig:cdfE2Erate} to the one in \ref{fig:cdfAccessRate}.	
	On the contrary, the end-to-end data-rates of the ad-hoc deployment outperform the one of the random deployment, 
	but are severely penalized by the backhaul links, 
	as shown by comparing the results in Figs. \ref{fig:cdfE2Erate} to the one in \ref{fig:cdfBackhaulSC}.
	%where the increased number of SCs served per mMIMO-BS significantly decrease the multiplexing capability of the mMIMO precoder.
	Thus, the reduced backhaul capacity of the mMIMO sBH ad-hoc deployment does not fully allow to exploit the potentially high data-rate achieved in the access. 
	This behavior suggests the need to optimize the splitting of resources between the two links. Indeed, a particularly important improvement in end-to-end rates would be achieved through the allocation of more resources to backhaul links.
	This is analyzed in the following section.
	
\end{itemize}

\subsection{Massive MIMO s-BH: Access and Backhaul Resource Allocation} \label{sec:ACandBHpartition}
	
\begin{figure}[!t]
	\centering
	\subfloat[$5$-th percentile]{{\includegraphics[width=0.90\columnwidth]{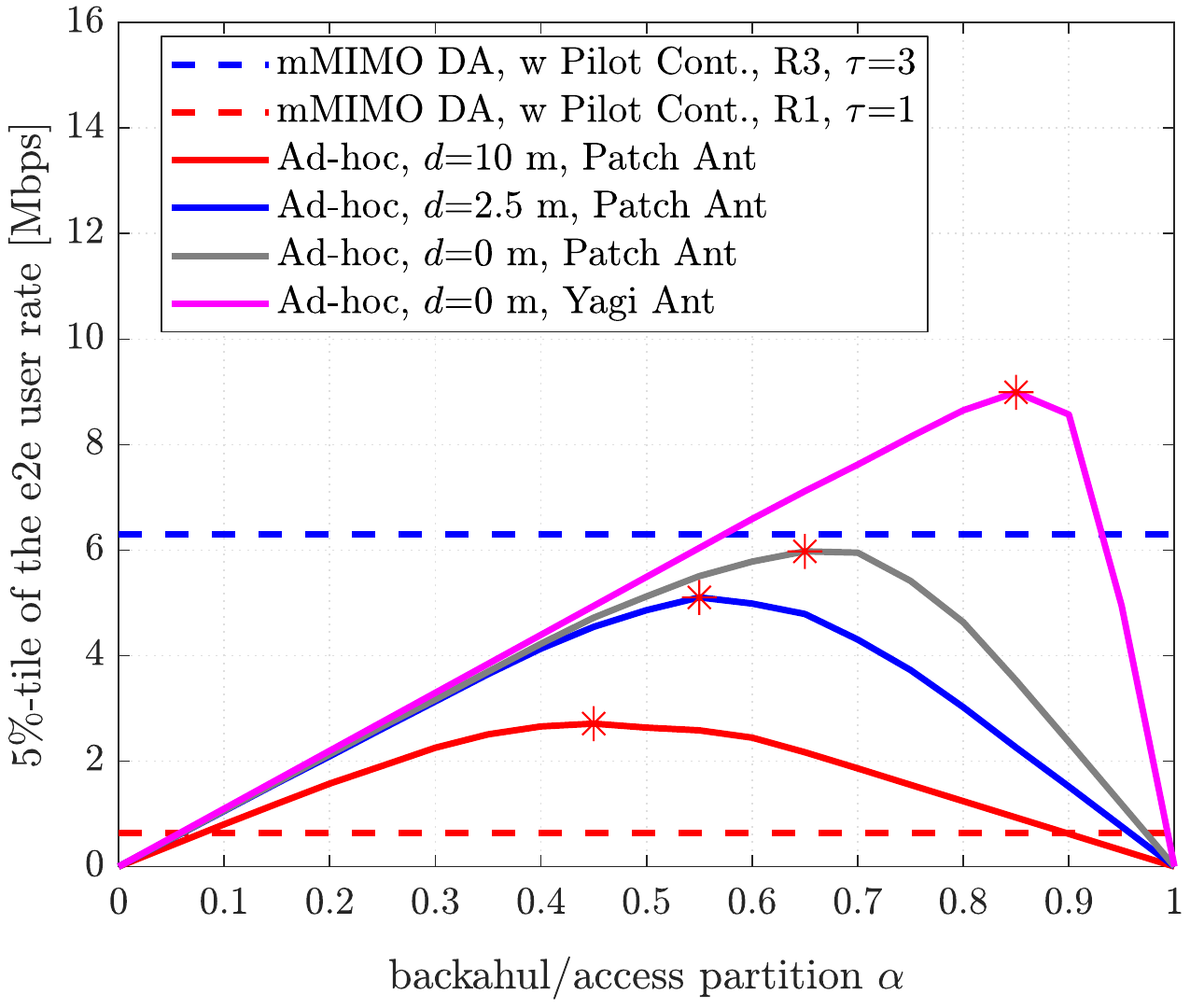}\label{fig:rateE2Efive}}}%
	%\qquad
	\hfill
	\subfloat[$50$-th percentile]{{\includegraphics[width=0.90\columnwidth]{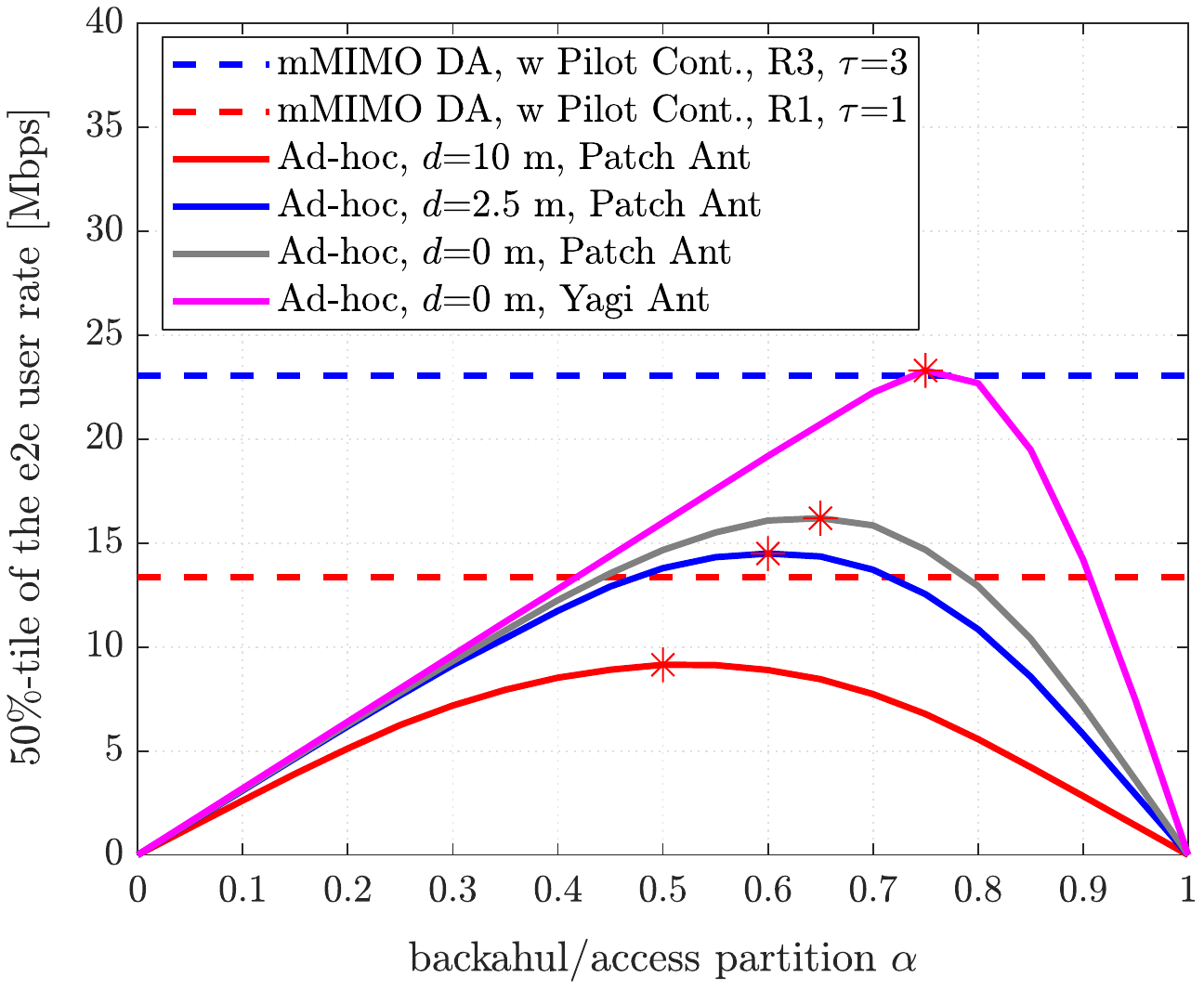}\label{fig:rateE2Efifty}}}%
	\caption{(a) $5$-th, and (b) $50$-th percentile of the end-to-end UE rates as a function of the fraction $\alpha$ of backhaul time-slots.}%
	\label{fig:resalloc} 
\end{figure}

In Fig. \ref{fig:resalloc}, 
we vary $\alpha$ in the range $0 \leq \alpha \leq 1$, 
and analyze the behavior of UEs rate at the $5$-th and $50$-th percentiles of the CDF. 
The configurations $\alpha=0$ and $\alpha=1$ entail that all the time-slots are allocated to the access and the backhaul, respectively. 
Therefore, the UE rates for these two values are equal to 0, 
since no resources are left for the other link. 
Moreover, the configuration $\alpha^*$ represents the value of $\alpha$ that maximizes the UE rate. 
For instance, with $d=0$ and Yagi antennas at the SCs, 
$\alpha^*$ is equal to 0.85 when looking at the $5$-th percentile.
Fig. \ref{fig:resalloc} brings the following insights:

%\begin{itemize}[leftmargin=*]
\begin{itemize}
	\item 
	By comparing the results between Fig. \ref{fig:rateE2Efive} and Fig. \ref{fig:rateE2Efifty}, 
	it is important to note that the optimal $\alpha$
	changes from 0.85 to 0.75.
	A tradeoff exists between $5$-th and $50$-th percentile performance, 
	and they cannot be optimized simultaneously.
	Assuming that the network uses $\alpha = 0.85$, 
	which is the optimal value for cell-edge UEs ($5$-th percentile of the CDF),
	the median UEs ($50$-th percentile of the CDF) can achieve an end-to-end rate of 19.5 Mbps, 
	which represents a $16\%$ reduction with respect to the maximum end-to-end rate achievable of 23.3 Mbps with $\alpha = 0.75$.
	\item 
	In Figs. \ref{fig:rateE2Efive} and \ref{fig:rateE2Efifty}, 
	we show with dashed lines the results of the mMIMO DA setup. 
	A properly designed mMIMO s-BH radio resource partitioning can improve the performance of the cell-edge UEs, 
	and keeps the same performance for the UEs at the median of the CDF with respect to mMIMO DA architecture, as shown in Figs. \ref{fig:rateE2Efive} and \ref{fig:rateE2Efifty}, respectively. 
	A more detailed comparison is further developed in the next section.
\end{itemize}

\subsection{Massive MIMO Architectures: s-BH vs. Direct Access} \label{sec:cdfRelayVsDA}

\begin{figure}[!t]
	\centering
	\includegraphics[width=0.90\columnwidth]{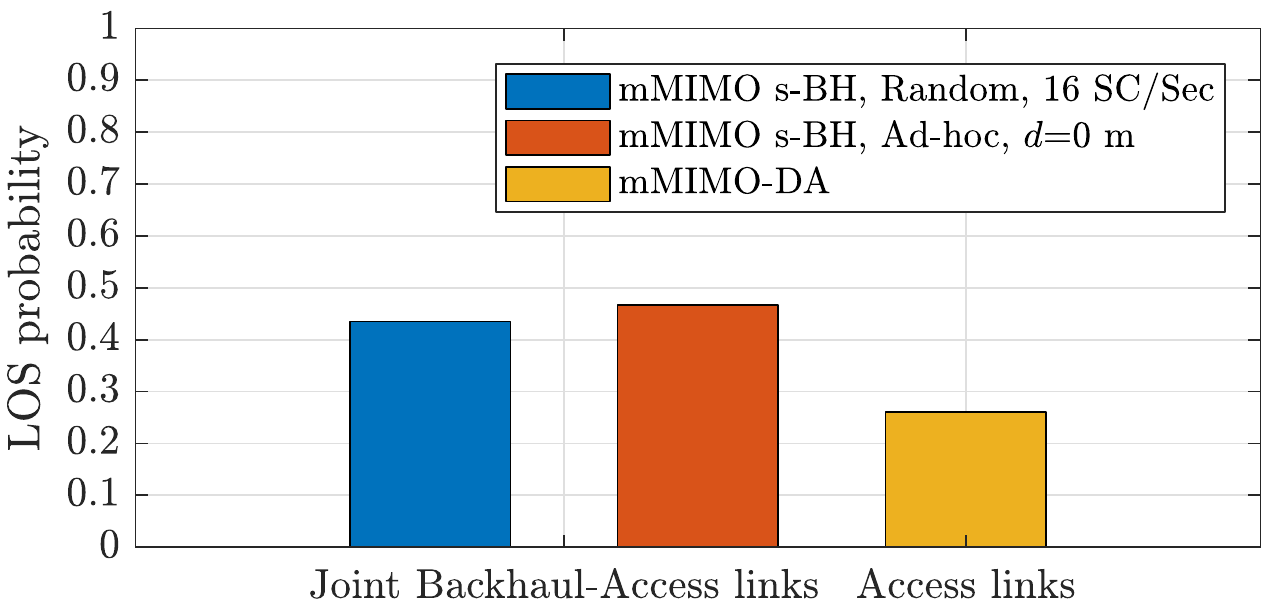}%
	\caption{LoS probability for joint backhaul-access links in s-BH network and for access links in DA network.}%
	\label{fig:losProbsBHvsDA}%
\end{figure}

First, we compare the joint probability of LoS-LoS channels in the backhaul-access legs for the s-BH network with the probability of LoS channel in the access links for the DA network.
As shown in Fig. \ref{fig:losProbsBHvsDA}, 
there are higher joint backhaul-access LoS probabilities with respect to the access LoS probability, $47\%$ and $25\%$, respectively. 
Those are the cases where the s-BH architecture can potentially improve the UE performance with respect to the DA architecture.

\begin{figure}[!t]
	\centering
	\includegraphics[width=0.90\columnwidth]{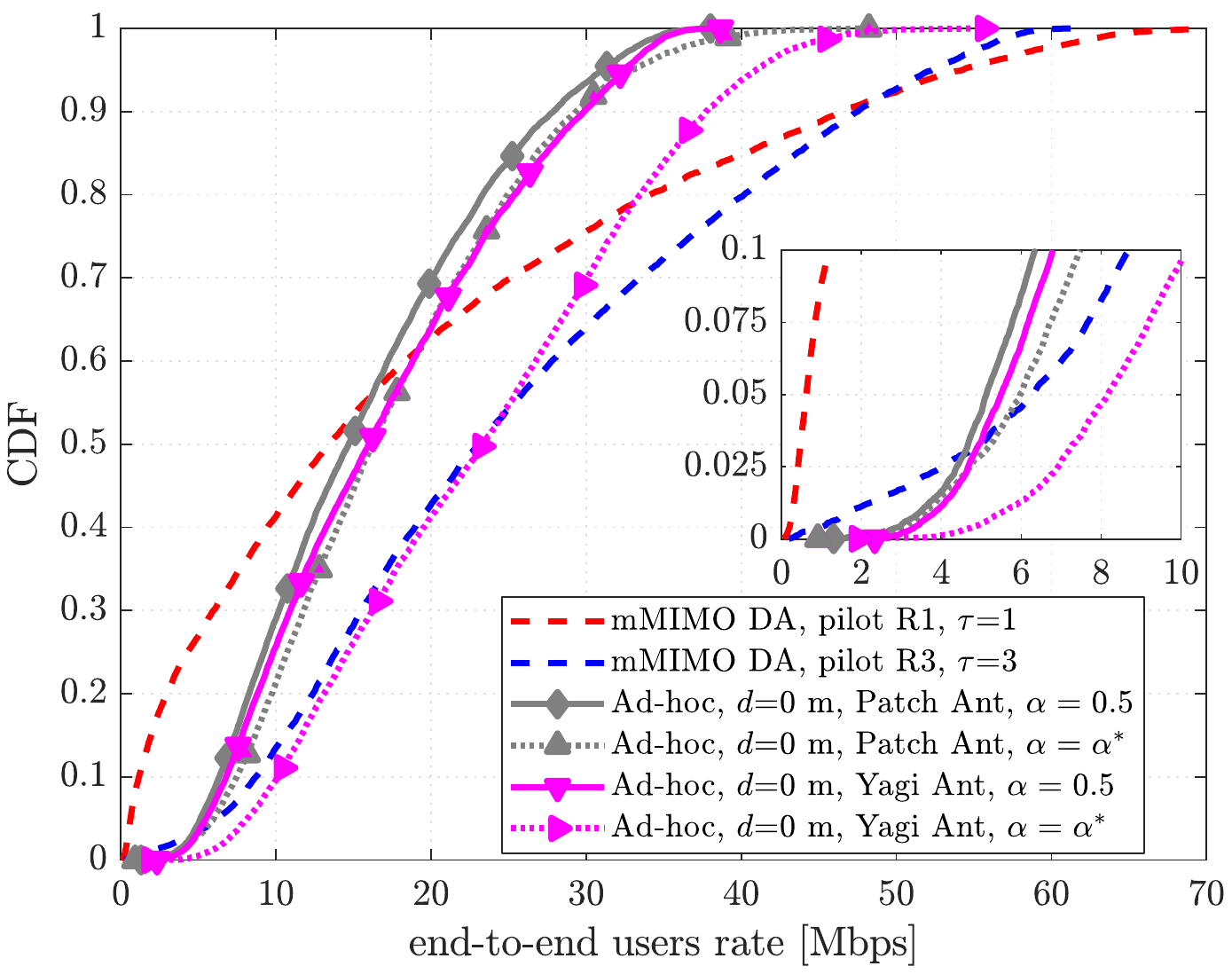}%
	\caption{Two types of curves are represented: ($i$) mMIMO DA with pilot reuse schemes 1 and 3; ($ii$) ad-hoc deployment of 16 SCs per sector for $\alpha=0.5$ and $\alpha=\alpha^{\ast}$, at which the $50$-th percentile of the end-to-end UE rate is maximized (as shown in Fig. \ref{fig:rateE2Efifty}).}
	\label{fig:cdfRelayVsDA}
\end{figure}

We also compare the end-to-end UE rates resulting from the 3GPP-based simulations of the mMIMO s-BH and mMIMO DA networks.
As shown in Fig. \ref{fig:cdfRelayVsDA}, 
the mMIMO s-BH network with the ad-hoc deployment of SCs provides better performance than the mMIMO DA network with pilot reuse 1 at the bottom of the CDF, 
i.e. below the $50$-th percentile. 
In fact, pilot contamination severely degrades the rates of UEs at the cell edge in the mMIMO DA setup with pilot reuse 1. 
On the other hand, in the s-BH network, due to the longer coherence time of the backhaul channel, $T_{BH}$, with respect to the system time-slot duration, $T$, 
there is no pilot contamination, 
and the UEs benefit from the proximity of SCs, 
which reduces the pathloss and improves the LoS propagation condition, 
as shown in Fig. \ref{fig:losProbsBHvsDA}.

However, by adopting the pilot reuse 3 in the mMIMO DA network, 
the pilot contamination effect reduces, 
%and the results show that the mMIMO DA network performance exceeds the one of the mMIMO s-BH network with $\alpha=0.5$, 
and the results show that the mMIMO DA performance exceeds the one of the mMIMO s-BH with $\alpha=0.5$, 
even if the pilot overhead ($\tau=3$ OFDM symbols) is 3 times larger with respect to pilot reuse 1. 
The mMIMO s-BH architecture provides the same performance as the mMIMO DA for the median UEs, 
only when the optimal partition $\alpha=\alpha^{\ast}$ is selected. 
Overall, mMIMO s-BH underperforms mMIMO DA above the median of the UE rates, 
and provides rate improvements for cell-edge UEs that amount to $30\%$ and a tenfold gain when adopting pilot reuse 3 and reuse 1, respectively.

\subsection{Asymptotic Data-rate Analysis} \label{subs:asymptoticRegime}

\begin{figure}[!t]
	\centering
	\subfloat[Asymptotic SC data-rate for backhaul links]{\includegraphics[width=0.90\columnwidth]{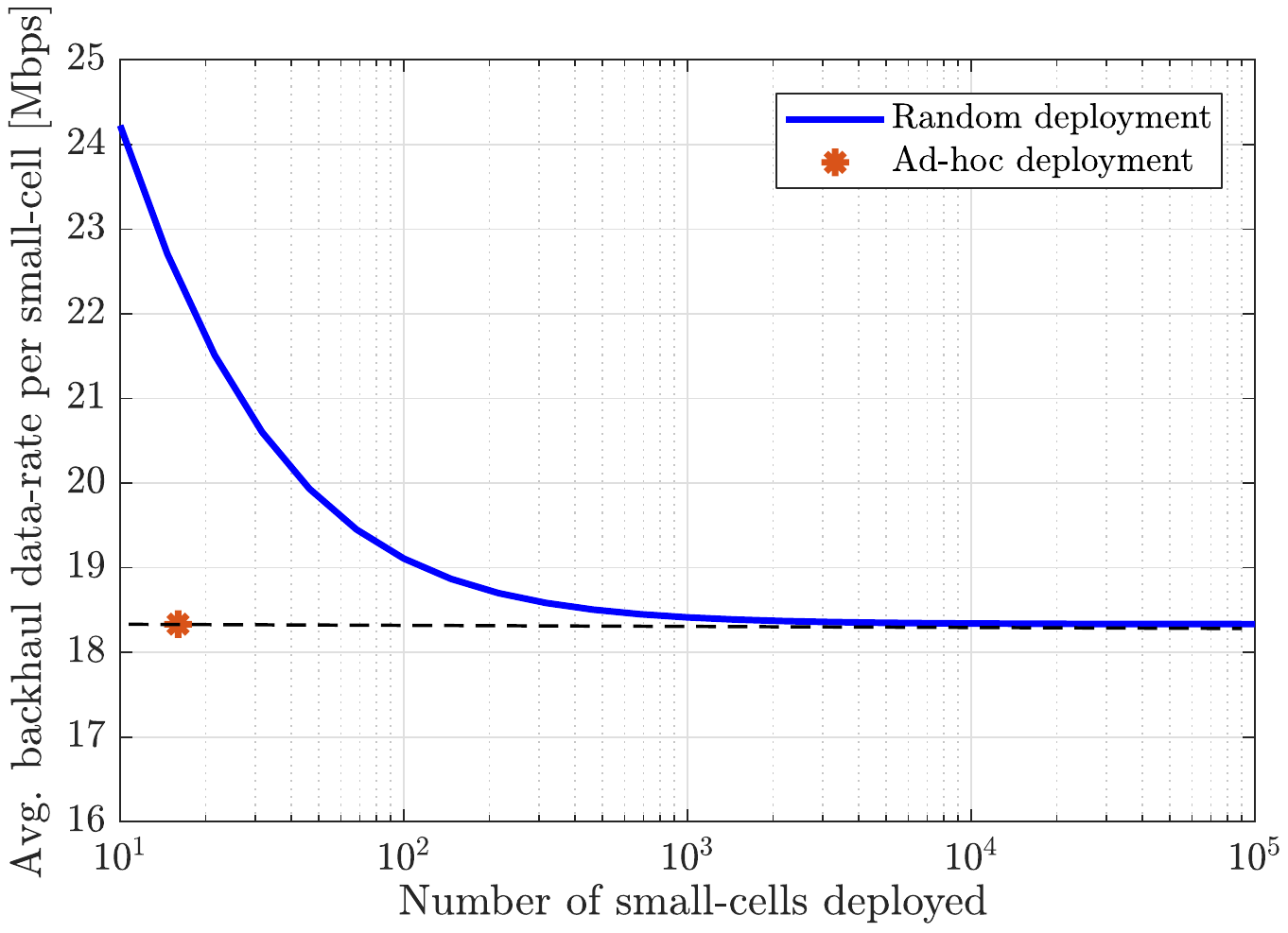}	\label{fig:asymptoticBackhaul}}%
	%\qquad
	\hfill
	\subfloat[Asymptotic UE data-rate for access links]{{\includegraphics[width=0.90\columnwidth]{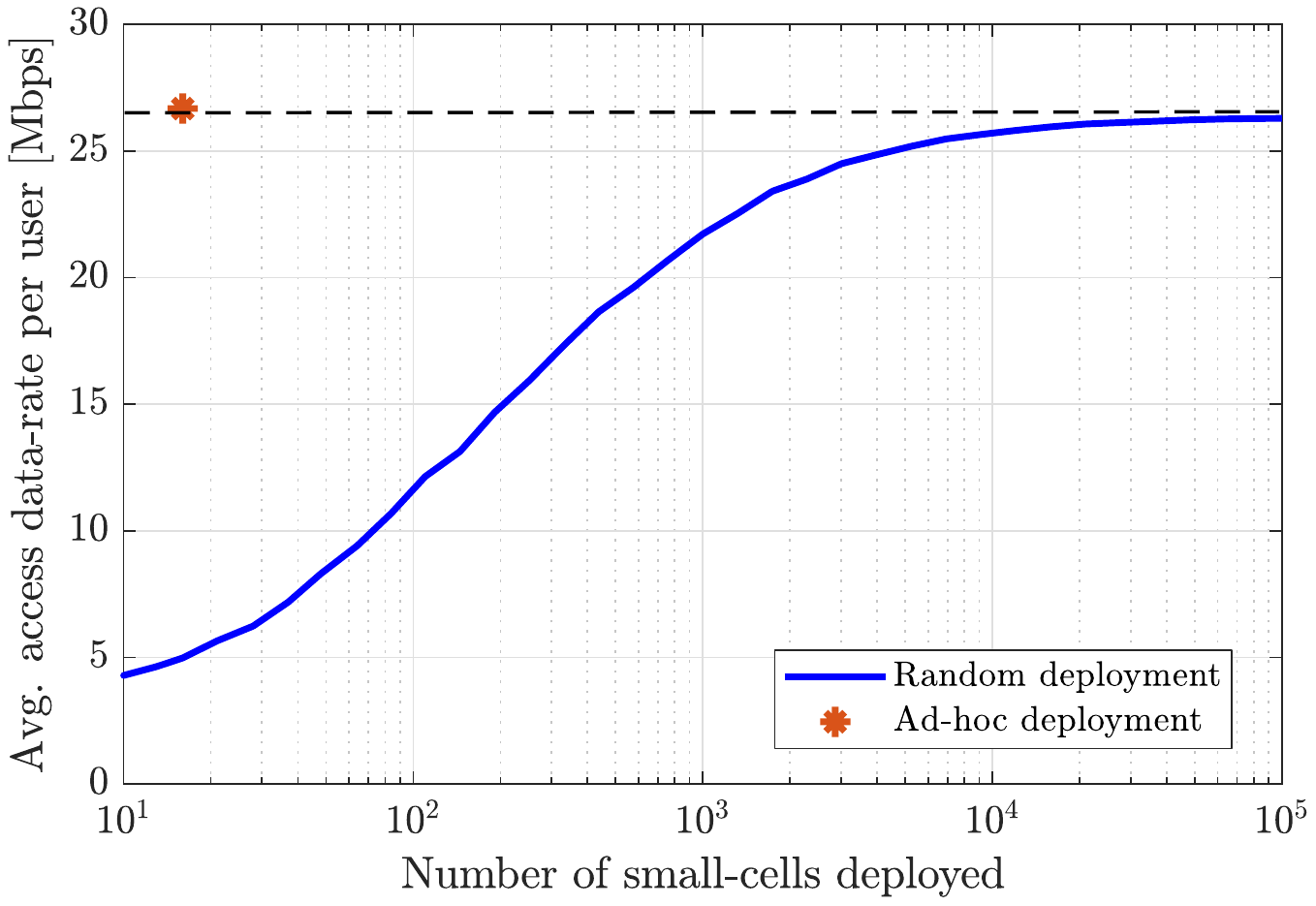} \label{fig:asymptoticAccess}}}%
	\caption{Asymptotic performance measures for backhaul and access links in s-BH network when random and ad-hoc deployments of SCs are considered.}%
	\label{fig:asymptoticCurves}%
\end{figure}

Figs. \ref{fig:asymptoticBackhaul} and \ref{fig:asymptoticAccess} show the convergence behavior of the backhaul and access data-rates for the random SCs deployment with respect to the results obtained with the ad-hoc deployment of 16 SCs, positioned at fixed distance $d=0$ with respect to the UEs. 
In both cases the results are obtained by numerical integration of \eqref{eq:avgBackhaul} and \eqref{eq:avgAccess}.
Fig. \ref{fig:asymptoticBackhaul} show the convergence results of the backhaul link. 
The backhaul data-rate of the random distribution (solid line) converges to the one of the ad-hoc (dashed line) when the number of SCs is 100 times larger than the number of ad-hoc SCs (denoted by the marker ``*'').
Fig. \ref{fig:asymptoticAccess} shows the convergence results of the access link. 
In this case, the data rate of the random (solid line) converges to the ad-hoc (dashed line), when the number of SCs is 1000 times larger than the number of SCs deployed in the ad-hoc case (marker ``*'').
As shown in subsection \ref{sec:ACandBHpartition}, the ad-hoc deployment is the one which maximizes the end-to-end data-rates of the two-hop communication.

The main takeaway is summarized by the possibility for the MNOs to explore the adoption of future dynamic SCs infrastructures \cite{8647997}. Indeed, instead of significantly over-provisioning the number of SCs, it may be beneficial to dynamically reposition only the active ones, trying to guarantee the same performance that are obtained with a very dense deployment of SCs.

\section{Conclusion and Future Works}\label{Sec:6}

In this paper, 
we studied the performance results for two 5G mMIMO architectures working at frequencies below 6 GHz: mMIMO s-BH and mMIMO DA. 
In the mMIMO s-BH architecture, 
we analyzed two different configurations: 
random deployment of SCs in the coverage area of the serving macro cell, 
and ad-hoc deployment of SCs in close proximity to each UE. 
The first takeaway of this study is that the random SCs distribution entails deploying thousands of SCs to achieve the access link performance upper bound.
On the other hand, the ad-hoc deployment benefits from the close proximity of the SCs to the UEs, 
and outperforms the random one for reasonable numbers of deployed SCs. 
However, the SC requires to know the UE position, and this is particularly complicated to realize due to mobility.
The second takeaway is that an ad-hoc SCs deployment supported by mMIMO s-BH provides rate improvements for cell-edge UEs that amount to 30\% and a tenfold gain as compared to mMIMO DA with pilot reuse 3 and reuse 1, respectively. 
This means that pilot contamination severely impacts the UEs performance at the cell-edge, and it is better to serve the UEs with the mMIMO s-BH architecture.
On the other hand, mMIMO DA outperforms s-BH above the median of the UE rates, meaning that when pilot contamination is less severe, and the LoS probability of the DA links improves, is better to serve the UEs with the mMIMO DA architecture.

Future works will focus on improving the capacity of the backhaul, 
which is currently the main limitation in the s-BH architecture at the sub-6 GHz frequencies.
At first, we aim to investigate how to improve the LoS probability in the backhaul links.
This improvement can be achieved by increasing the altitude of the s-BH SCs, for example by means of drone-BSs hovering above the UE locations.
Secondly, we aim to increase the bandwidth and limit the interference of the backhaul links.
This can be realized by using the mmWave frequencies which can simultaneously guarantee the availability of abundant spectrum in the mmWave bands, and partially avoid the inter-cell interference issue, due to the high directionality of the transmission at those carrier frequencies.

% use section* for acknowledgment
\section*{Acknowledgment}
The authors would like to thank Dr. Jacek Kibilda for his valuable suggestions and helpful comments.

\ifCLASSOPTIONcaptionsoff
\newpage
\fi

\bibliographystyle{IEEEtran}
\bibliography{Strings_Gio,mycollection}

% Generated by IEEEtran.bst, version: 1.12 (2007/01/11)
\begin{thebibliography}{10}
\providecommand{\url}[1]{#1}
\csname url@samestyle\endcsname
\providecommand{\newblock}{\relax}
\providecommand{\bibinfo}[2]{#2}
\providecommand{\BIBentrySTDinterwordspacing}{\spaceskip=0pt\relax}
\providecommand{\BIBentryALTinterwordstretchfactor}{4}
\providecommand{\BIBentryALTinterwordspacing}{\spaceskip=\fontdimen2\font plus
\BIBentryALTinterwordstretchfactor\fontdimen3\font minus
  \fontdimen4\font\relax}
\providecommand{\BIBforeignlanguage}[2]{{%
\expandafter\ifx\csname l@#1\endcsname\relax
\typeout{** WARNING: IEEEtran.bst: No hyphenation pattern has been}%
\typeout{** loaded for the language `#1'. Using the pattern for}%
\typeout{** the default language instead.}%
\else
\language=\csname l@#1\endcsname
\fi
#2}}
\providecommand{\BIBdecl}{\relax}
\BIBdecl

\bibitem{8647638}
A.~{Bonfante}, L.~{Galati Giordano}, D.~{Lopez-Perez}, A.~{Garcia-Rodriguez},
  G.~{Geraci}, P.~{Baracca}, M.~M. {Butt}, M.~{Dzaferagic}, and N.~{Marchetti},
  ``Performance of massive {MIMO} self-backhauling for ultra-dense small cell
  deployments,'' in \emph{Proc. IEEE Global Commun. Conf. (Globecom)}, Dec.
  2018, pp. 1--7.

\bibitem{7169508}
A.~Gupta and R.~K. Jha, ``A survey of {5G} network: {A}rchitecture and emerging
  technologies,'' \emph{IEEE Access}, vol.~3, pp. 1206--1232, Jul. 2015.

\bibitem{6375940}
F.~Rusek, D.~Persson, B.~K. Lau, E.~G. Larsson, T.~L. Marzetta, O.~Edfors, and
  F.~Tufvesson, ``Scaling up {MIMO}: {O}pportunities and challenges with very
  large arrays,'' \emph{IEEE Signal Processing Mag.}, vol.~30, no.~1, pp.
  40--60, Jan. 2013.

\bibitem{7126919}
D.~L{\'{o}}pez{-}P{\'{e}}rez, M.~Ding, H.~Claussen, and A.~H. Jafari, ``Towards
  1 {Gbps/UE} in cellular systems: {U}nderstanding ultra-dense small cell
  deployments,'' \emph{IEEE Communications Surveys Tutorials}, vol.~17, no.~4,
  pp. 2078--2101, Fourthquarter 2015.

\bibitem{7306534}
U.~Siddique, H.~Tabassum, E.~Hossain, and D.~I. Kim, ``Wireless backhauling of
  {5G} small cells: challenges and solution approaches,'' \emph{IEEE Wireless
  Communications}, vol.~22, no.~5, pp. 22--31, Oct. 2015.

\bibitem{7306536}
N.~Wang, E.~Hossain, and V.~K. Bhargava, ``Backhauling {5G} small cells: A
  radio resource management perspective,'' \emph{IEEE Wireless Communications},
  vol.~22, no.~5, pp. 41--49, Oct. 2015.

\bibitem{TR_38.874}
{3GPP TR 38.874}, ``{Study on integrated access and backhaul},'' {Technical
  Report (TR)}, May 2018.

\bibitem{7974805}
M.~N. Kulkarni, J.~G. Andrews, and A.~Ghosh, ``Performance of dynamic and
  static {TDD} in self-backhauled millimeter wave cellular networks,''
  \emph{IEEE Trans. Wireless Commun.}, vol.~16, no.~10, pp. 6460--6478, Oct.
  2017.

\bibitem{Singh2015}
S.~Singh, M.~N. Kulkarni, A.~Ghosh, and J.~G. Andrews, ``Tractable model for
  rate in self-backhauled millimeter wave cellular networks,'' \emph{IEEE J.
  Sel. Areas Commun.}, vol.~33, no.~10, pp. 2196--2211, Oct. 2015.

\bibitem{7962657}
R.~Gupta and S.~Kalyanasundaram, ``Resource allocation for self-backhauled
  networks with half-duplex small cells,'' in \emph{Proc. IEEE Int. Conf. on
  Comm. (ICC)}, May 2017, pp. 198--204.

\bibitem{Nguyen2016}
T.~M. Nguyen, A.~Yadav, W.~Ajib, and C.~Assi, ``Resource allocation in two-tier
  wireless backhaul heterogeneous networks,'' \emph{IEEE Trans. Wireless
  Commun.}, vol.~15, no.~10, pp. 6690--6704, Oct. 2016.

\bibitem{7817893}
A.~Sharma, R.~K. Ganti, and J.~K. Milleth, ``Joint backhaul-access analysis of
  full duplex self-backhauling heterogeneous networks,'' \emph{IEEE Trans.
  Wireless Commun.}, vol.~16, no.~3, pp. 1727--1740, Mar. 2017.

\bibitem{7177124}
B.~Li, D.~Zhu, and P.~Liang, ``Small cell in-band wireless backhaul in massive
  {MIMO} systems: A cooperation of next-generation techniques,'' \emph{IEEE
  Trans. Wireless Commun.}, vol.~14, no.~12, pp. 7057--7069, Dec. 2015.

\bibitem{Tabassum}
H.~Tabassum, A.~H. Sakr, and E.~Hossain, ``Analysis of massive {MIMO}-enabled
  downlink wireless backhauling for full-duplex small cells,'' \emph{IEEE
  Trans. Commun.}, vol.~64, no.~6, pp. 2354--2369, Jun. 2016.

\bibitem{7445888_GioBackhaul}
H.~H. Yang, G.~Geraci, and T.~Q.~S. Quek, ``Energy-efficient design of {MIMO}
  heterogeneous networks with wireless backhaul,'' \emph{IEEE Trans. Wireless
  Commun.}, vol.~15, no.~7, pp. 4914--4927, Jul. 2016.

\bibitem{8241817}
M.~Feng, S.~Mao, and T.~Jiang, ``Joint frame design, resource allocation and
  user association for massive {MIMO} heterogeneous networks with wireless
  backhaul,'' \emph{IEEE Trans. Wireless Commun.}, vol.~17, no.~3, pp.
  1937--1950, Mar. 2018.

\bibitem{Lim}
Y.~G. Lim, C.~B. Chae, and G.~Caire, ``Performance analysis of massive {MIMO}
  for cell-boundary users,'' \emph{IEEE Trans. Wireless Commun.}, vol.~14,
  no.~12, pp. 6827--6842, Dec. 2015.

\bibitem{3gpp.36.814}
{3GPP 36.814}, ``{Further advancements for E-UTRA physical layer aspects},''
  {Technical Report (TR)}, Mar. 2017.

\bibitem{Kelif2010}
J.~M. Kelif, M.~Coupechoux, and P.~Godlewski, ``A fluid model for performance
  analysis in cellular networks,'' \emph{EURASIP Journ. Wireless Commun.
  Network.}, vol. 2010, no. 435189, Aug. 2010.

\bibitem{6702841}
M.~Minelli, M.~Ma, M.~Coupechoux, J.~M. Kelif, M.~Sigelle, and P.~Godlewski,
  ``Optimal relay placement in cellular networks,'' \emph{IEEE Trans. Wireless
  Commun.}, vol.~13, no.~2, pp. 998--1009, Feb. 2014.

\bibitem{5288507}
V.~Chandrasekhar and J.~G. Andrews, ``Spectrum allocation in tiered cellular
  networks,'' \emph{IEEE Trans. Commun.}, vol.~57, no.~10, pp. 3059--3068, Oct.
  2009.

\bibitem{5165314}
------, ``Uplink capacity and interference avoidance for two-tier femtocell
  networks,'' \emph{IEEE Trans. Wireless Commun.}, vol.~8, no.~7, pp.
  3498--3509, Jul. 2009.

\bibitem{8403635}
C.~{Liu}, M.~{Ding}, C.~{Ma}, Q.~{Li}, Z.~{Lin}, and Y.~{Liang}, ``Performance
  analysis for practical unmanned aerial vehicle networks with {LoS/NLoS}
  transmissions,'' in \emph{Proc. IEEE Int. Conf. on Comm. Workshops (ICC
  Workshops)}, May 2018, pp. 1--6.

\bibitem{6241389}
H.~Huh, G.~Caire, H.~C. Papadopoulos, and S.~A. Ramprashad, ``Achieving
  {``Massive MIMO''} spectral efficiency with a not-so-large number of
  antennas,'' \emph{IEEE Trans. Wireless Commun.}, vol.~11, no.~9, pp.
  3226--3239, Sep. 2012.

\bibitem{3gpp.25.996}
{3GPP 25.996}, ``{{Spatial channel model for Multiple Input Multiple Output
  (MIMO) simulations (Release 14)}},'' {Technical Report (TR)}, Mar. 2017.

\bibitem{marzetta2016fundamentals}
T.~L. Marzetta, E.~G. Larsson, H.~Yang, and H.~Q. Ngo, ``Fundamentals of
  massive {MIMO},'' \emph{Cambridge University Press}, 2016.

\bibitem{1237141}
H.~Shin and J.~H. Lee, ``Capacity of multiple-antenna fading channels: spatial
  fading correlation, double scattering, and keyhole,'' \emph{IEEE Trans. Inf.
  Theory}, vol.~49, no.~10, pp. 2636--2647, Oct. 2003.

\bibitem{4277071}
A.~Forenza, D.~J. Love, and R.~W. Heath, ``Simplified spatial correlation
  models for clustered {MIMO} channels with different array configurations,''
  \emph{IEEE Trans. Veh. Technol.}, vol.~56, no.~4, pp. 1924--1934, Jul. 2007.

\bibitem{molisch2010wireless}
A.~Molisch, ``Wireless communications,'' \emph{John Wiley \& Sons}, 2010.

\bibitem{656151}
R.~B. Ertel, P.~Cardieri, K.~W. Sowerby, T.~S. Rappaport, and J.~H. Reed,
  ``Overview of spatial channel models for antenna array communication
  systems,'' \emph{IEEE Personal Communications}, vol.~5, no.~1, pp. 10--22,
  Feb. 1998.

\bibitem{Zhu2016}
X.~Zhu, Z.~Wang, C.~Qian, L.~Dai, J.~Chen, S.~Chen, and L.~Hanzo, ``Soft pilot
  reuse and multicell block diagonalization precoding for massive {MIMO}
  systems,'' \emph{IEEE Trans. Veh. Technol.}, vol.~65, no.~5, pp. 3285--3298,
  May 2016.

\bibitem{6415397}
H.~Yin, D.~Gesbert, M.~Filippou, and Y.~Liu, ``A coordinated approach to
  channel estimation in large-scale multiple-antenna systems,'' \emph{IEEE J.
  Sel. Areas Commun.}, vol.~31, no.~2, pp. 264--273, Feb. 2013.

\bibitem{Galati1712}
L.~{Galati Giordano}, L.~Campanalonga, D.~{L{\'o}pez-P{\'e}rez},
  A.~Garcia-Rodriguez, G.~Geraci, P.~Baracca, and M.~Magarini, ``Uplink
  sounding reference signal coordination to combat pilot contamination in {5G
  massive MIMO},'' in \emph{Proc. IEEE Wireless Commun. Networking Conference
  (WCNC)}, Apr. 2018, pp. 1--6.

\bibitem{1261332}
Q.~H. {Spencer}, A.~L. {Swindlehurst}, and M.~{Haardt}, ``Zero-forcing methods
  for downlink spatial multiplexing in multiuser {MIMO} channels,'' \emph{IEEE
  Trans. Signal Process.}, vol.~52, no.~2, pp. 461--471, Feb. 2004.

\bibitem{massivemimobook}
E.~Bj\"{o}rnson, J.~Hoydis, and L.~Sanguinetti, ``Massive {MIMO} networks:
  {Spectral}, energy, and hardware efficiency,'' \emph{Foundations and Trends
  in Signal Processing}, vol.~11, no. 3-4, pp. 154--655, 2017.

\bibitem{haenggi_2012}
M.~Haenggi, ``Stochastic geometry for wireless networks,'' \emph{Cambridge
  University Press}, 2012.

\bibitem{6775036}
C.~Li, J.~Zhang, and K.~B. Letaief, ``Throughput and energy efficiency analysis
  of small cell networks with multi-antenna base stations,'' \emph{IEEE Trans.
  Wireless Commun.}, vol.~13, no.~5, pp. 2505--2517, May 2014.

\bibitem{6205422}
S.~Lee and K.~Huang, ``Coverage and economy of cellular networks with many base
  stations,'' \emph{IEEE Commun. Letters}, vol.~16, no.~7, pp. 1038--1040, Jul.
  2012.

\bibitem{illian2008statistical}
J.~Illian, A.~Penttinen, H.~Stoyan, and D.~Stoyan, ``Statistical analysis and
  modelling of spatial point patterns,'' \emph{John Wiley \& Sons}, vol.~70,
  2008.

\bibitem{5621983}
J.~G. Andrews, R.~K. Ganti, M.~Haenggi, N.~Jindal, and S.~Weber, ``A primer on
  spatial modeling and analysis in wireless networks,'' \emph{IEEE Comms.
  Mag.}, vol.~48, no.~11, pp. 156--163, Nov. 2010.

\bibitem{7335646}
M.~Ding, P.~Wang, D.~{L{\'o}pez-P{\'e}rez}, G.~Mao, and Z.~Lin, ``Performance
  impact of {LoS} and {NLoS} transmissions in dense cellular networks,''
  \emph{IEEE Trans. Wireless Commun.}, vol.~15, no.~3, pp. 2365--2380, Mar.
  2016.

\bibitem{7248759}
C.~Galiotto, N.~K. Pratas, N.~Marchetti, and L.~Doyle, ``A stochastic geometry
  framework for {LoS}/{NLoS} propagation in dense small cell networks,'' in
  \emph{Proc. IEEE Int. Conf. on Comm. (ICC)}, Jun. 2015, pp. 2851--2856.

\bibitem{6497002}
S.~Singh, H.~S. Dhillon, and J.~G. Andrews, ``Offloading in heterogeneous
  networks: Modeling, analysis, and design insights,'' \emph{IEEE Trans.
  Wireless Commun.}, vol.~12, no.~5, pp. 2484--2497, May 2013.

\bibitem{8515110}
G.~{George}, A.~{Lozano}, and M.~{Haenggi}, ``Distribution of the number of
  users per base station in cellular networks,'' \emph{IEEE Wireless
  Communications Letters}, vol.~8, no.~2, pp. 520--523, Apr. 2019.

\bibitem{8647997}
J.~{Kibilda} and G.~{de Veciana}, ``Dynamic network densification: Overcoming
  spatio-temporal variability in wireless traffic,'' in \emph{Proc. IEEE Int.
  Conf. on Comm. (ICC)}, Dec. 2018, pp. 1--6.

\end{thebibliography}

\appendices

\section{} \label{appendix}

In the following, we derive the expression used to evaluate the rate coverage probability.
The probability that the access SIR is greater than a threshold  $\gamma_a = 2^{R_{th}\mu_{l}}-1$, which depends on the minimum target rate $R_{th}$, is expressed as \cite{7248759}

\begin{multline}
\Pr \left[{\mathrm{SIR}^{\mathrm{A}}}(x)>\gamma_a \right]  = \Pr \left[ \frac{P_{l}^{\mathrm{dl}} G_{b} |g|^2 \beta^{\mathrm{L}}(x)}{I_{{\mathrm {agg}}}}>\gamma_a \right] \overset{(a)}{=} \\ \mathscr{L}_{I_{{\mathrm {agg}}}}
\left(\frac{\gamma_a}{P_{l}^{\mathrm{dl}} G_{b} \beta^{\mathrm{L}}(x)}\right),
\end{multline}
where $(a)$ follows from \cite[eq. (54)]{7335646} neglecting the thermal noise as the propagation in sub-6 GHz bands is interference-limited and $\mathscr{L}_{I_{{\mathrm {agg}}}}(s)$ represents the Laplace transform of $I_{{\mathrm {agg}}}$, and is defined according to \cite{7335646,7248759} as follows
\begin{multline} 
\mathscr{L}_{I_{{\mathrm {agg}}}}(s)=\exp \left(-2\pi \tilde{\lambda_{b}}\int _{x}^{+\infty }
\frac{{\mathrm {Pr}}^{{\mathrm {L}}}\left(u\right)u}{1+\left(s P_{l}^{\mathrm{dl}} G_{b} \beta^{\mathrm{L}}\left(u\right) \right)^{-1}}du\right) \\
\times \exp \left(-2\pi \tilde{\lambda_{b}}\int _{x_1}^{+\infty }\frac{\left[1-{\mathrm {Pr}}^{{\mathrm {L}}}\left(u\right)\right]u}{1+\left(s P_{l}^{\mathrm{dl}} G_{b} \beta^{\mathrm{NL}}\left(u\right) \right)^{-1}}du\right),
\end{multline}
where $x_1=\left(\frac{A^{\mathrm{NL}}}{A^{\mathrm{L}}} \right)^{{\eta^{\mathrm{NL}}}^{-1}} x^{\frac{\eta^{\mathrm{L}}}{\eta^{\mathrm{NL}}}}$ \cite{7248759}.

% biography section
\begin{IEEEbiography}[{\includegraphics[width=1in,height=1.25in,keepaspectratio]{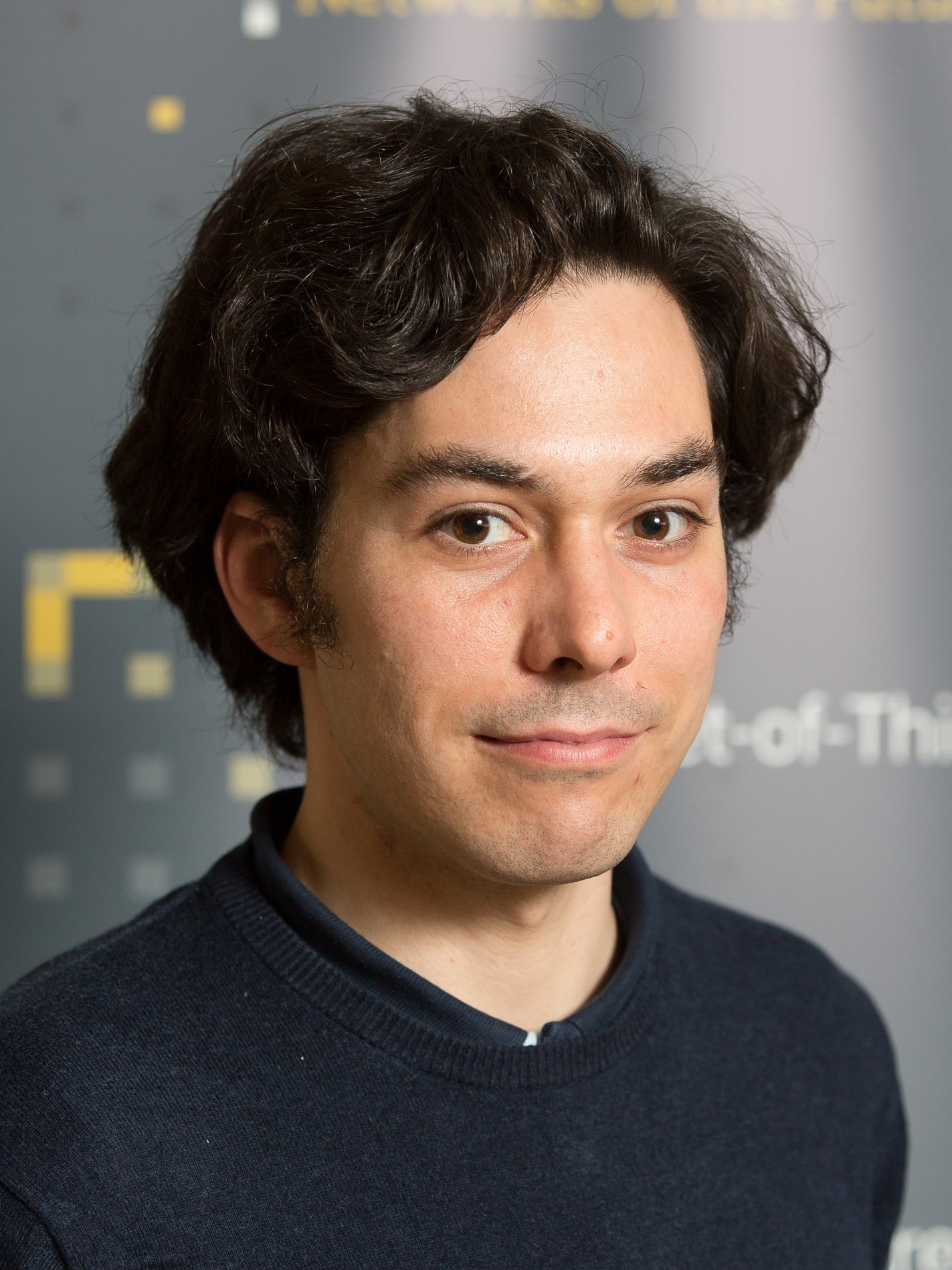}}]%
	{Andrea Bonfante} (S'18) is currently Ph.D. student at Trinity College Dublin supported by IRC Enterprise Partnership Scheme and co-founded by Nokia Bell Labs--Ireland. He received the B.Sc. and M.Sc. degrees in Telecommunications Engineering from Politecnico di Milano, Italy, in 2009 and 2012, respectively. From 2015 to 2016 he worked at Nokia Bell Labs, Dublin, Ireland, and previously at Azcom Technology, Milan, Italy, contributing to several industrial and research projects focusing on various wireless communication systems, such as massive MIMO, small cell base stations and relay systems. His ongoing studies focus on the network and system-level aspects of emerging wireless technologies, including cost-effective solutions for network deployment and optimal resource allocation for multi-hop networks. He regularly serves as reviewer for several Journals of IEEE, including \textsc{IEEE Transactions on Wireless Communications}, \textsc{IEEE Communications Letters} and \textsc{IEEE Access}. He served as TPC Member of numerous conferences and workshops including IEEE ICC'19, IEEE VTC'19 and EuCNC'19.
\end{IEEEbiography}

\begin{IEEEbiography}[{\includegraphics[width=1in,height=1.25in,keepaspectratio]{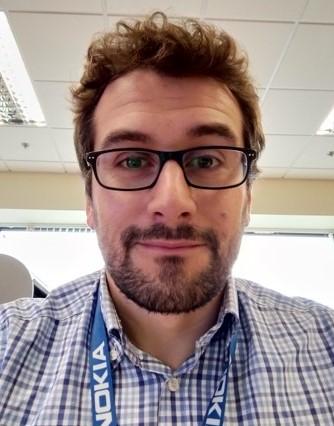}}]%
	{Lorenzo Galati Giordano} (M'15) is Member of Technical Staff at Nokia Bell Labs Ireland since 2015. Lorenzo received the M.Sc. and the Ph.D. degrees in wireless communication from Politecnico di Milano, Italy, in 2005 and 2010, respectively, and the master's degree in Innovation Management from IlSole24Ore Business School, Italy, in 2014. He was also Marie-Curie Short Term Fellow at University of Bedfordshire (UK) in 2008, researcher associate with the Italian National Research Council in 2010 and R\&D Engineer for Azcom Technology, an Italian SME, from 2010 to 2014. Lorenzo has more than 10 years of academical and industrial research experience on wireless communication systems and protocols, holds commercial patents and publications in prestigious books, IEEE journals and conferences. In addition, he has been Chair of IEEE WCNC Co-HetNEt 2012 workshop, editor of the book Vehicular Technologies -- Deployment and Applications by Intech, guest editor for \textsc{IEEE Access} special sections Networks of Unmanned Aerial Vehicles, TPC Member and reviewer for IEEE conferences and journals. Lorenzo is also very active in organizing tutorials and industrial seminars at major IEEE conferences, like ``Drone Base Stations: Opportunities and Challenges Towards a Truly ``Wireless'' Wireless Network'', which won the Most Attended Industry Program award at IEEE Globecom 2017. During the past years, Lorenzo contributed to the Nokia F-Cell project, an innovative self-powered and auto-connected drone deployed small cell served by massive MIMO wireless backhaul, which received the prestigious CTIA Emerging Technology 2016 Award. Lorenzo's current focus is on future indoor networks and next generation Wi-Fi technologies, an area where he is contributing with pioneering works on large antenna arrays solutions for the unlicensed spectrum.
\end{IEEEbiography}

\begin{IEEEbiography}[{\includegraphics[width=1in,height=1.25in,keepaspectratio]{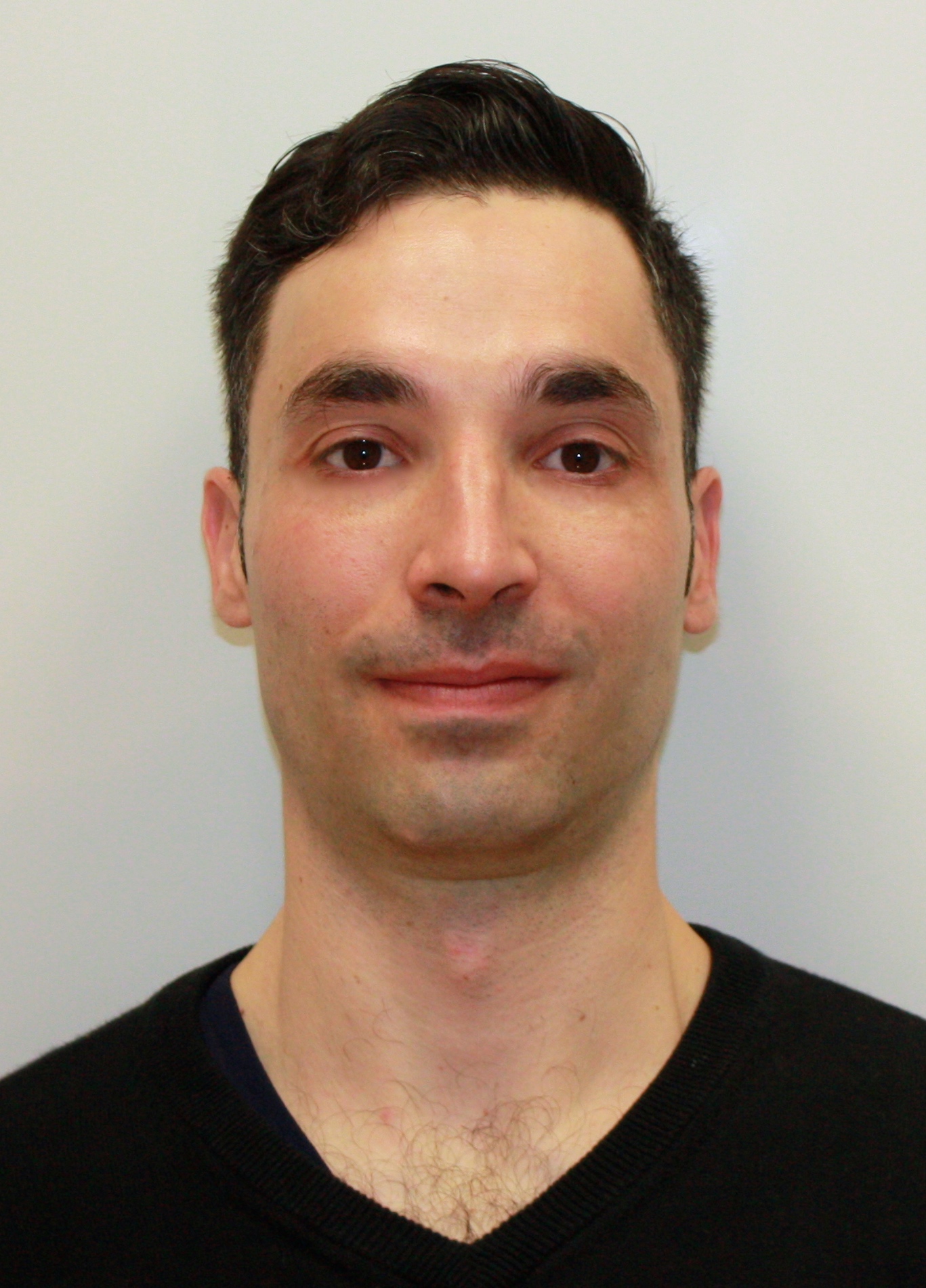}}]%
 	{David L\'{o}pez-P\'{e}rez} (M'12--SM'17)  is a Senior Research Scientist and a Distinguished Member of Technical Staff (DMTS) at Nokia Bell Laboratories, and his main research interests are on small cells, ultra-dense networks and unlicensed spectrum technologies, where he has pioneered work on LTE and Wi-Fi interworking. David is currently working on massive MIMO, future indoor networks and the next generation of Wi-Fi technology, IEEE802.11be. David has authored more than 135 research articles, holds over 49 patents applications, and has received a number of prestigious awards. He is an editor of \textsc{IEEE Transactions on Wireless Communications}.
\end{IEEEbiography}

\begin{IEEEbiography}[{\includegraphics[width=1in,height=1.25in,keepaspectratio]{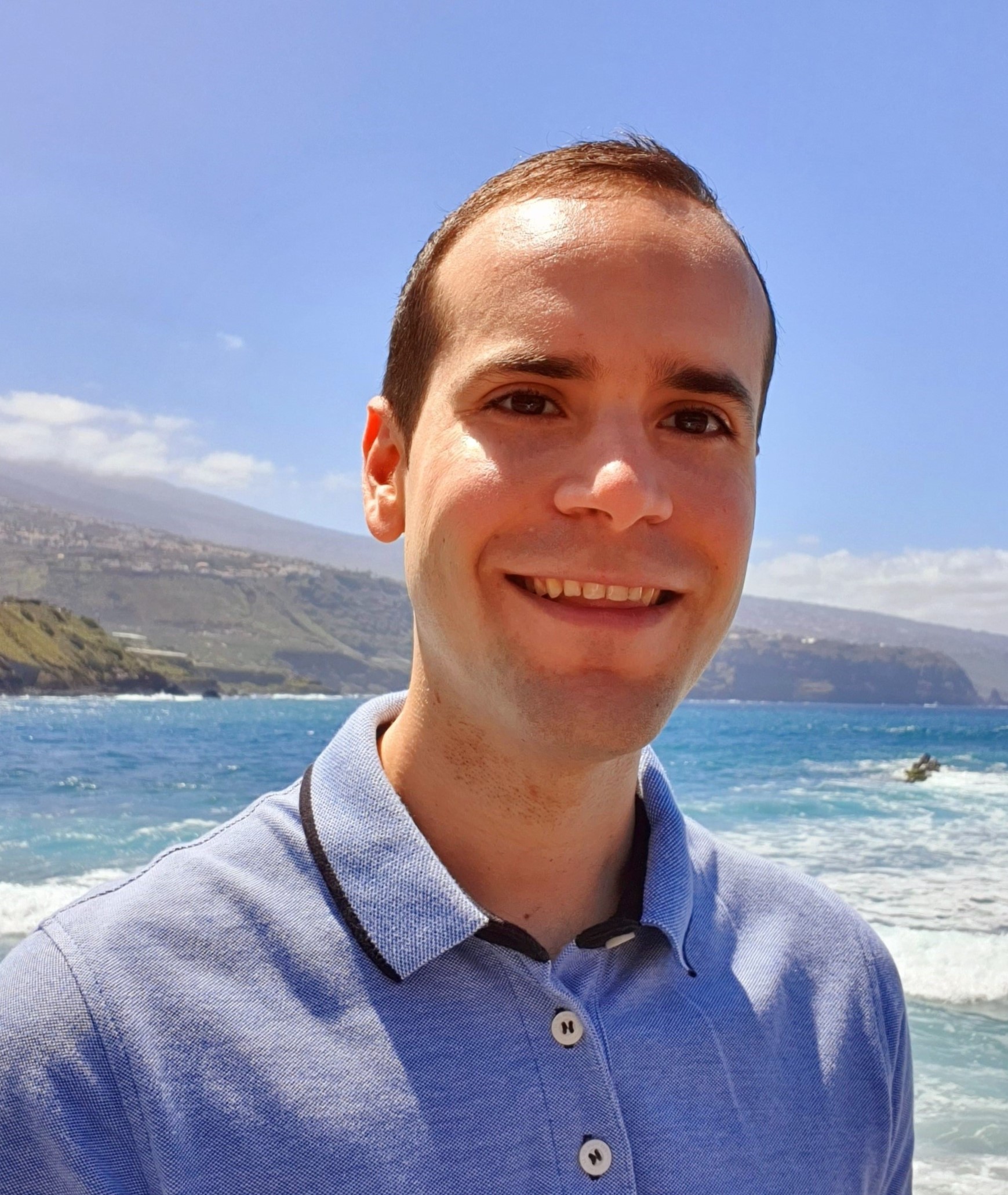}}]%
	{Adrian Garcia-Rodriguez} (S'13--M'17) is a Research Scientist in Nokia Bell Labs (Ireland), where he focuses on the design of next-generation 802.11 technologies and UAV communications. He joined Bell Labs in 2016, after receiving the Ph.D. degree in Electrical and Electronic Engineering from University College London (U.K.). Previously, he held research positions in the research institute for technological development and Communication Innovation (IDeTIC) at the University of Las Palmas de Gran Canaria (Spain) between 2010--2012, and in the RF group of Nokia Bell Labs (Ireland) in 2015. Adrian is co-inventor of fourteen filed patent families and co-author of 40+ technical publications in the areas of wireless communications and networking. On these topics, he frequently delivers technical tutorials (IEEE WCNC'18, IEEE ICC'18, IEEE PIMRC'18, and IEEE Globecom'18), organizes workshops and special sessions (IEEE Globecom'17, Asilomar'18, and IEEE ICC'19), and participates in industrial seminars (IEEE Globecom'18, and IEEE ICC'19). Adrian was named an Exemplary Reviewer for \textsc{IEEE Communications Letters} in 2016, and both \textsc{IEEE Transactions on Wireless Communications} and \textsc{IEEE Transactions on Communications} in 2017.
\end{IEEEbiography}

\begin{IEEEbiography}[{\includegraphics[width=1in,height=1.25in,keepaspectratio]{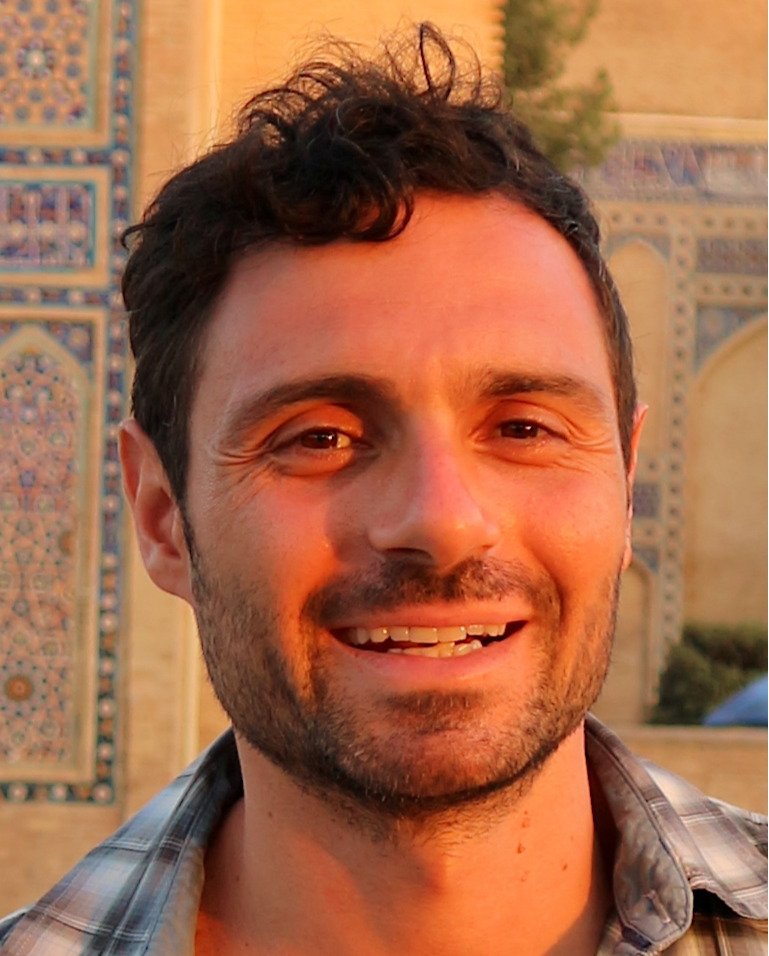}}]%
	{Giovanni Geraci} (S'11--M'14) is an Assistant Professor and Junior Leader Fellow at Universitat Pompeu Fabra in Barcelona (Spain). He earned a Ph.D. degree from the University of New South Wales (Australia) in 2014. He gained industrial innovation experience at Nokia Bell Labs (Ireland), where he was a Research Scientist in 2016-2018. His background also features appointments at the Singapore University of Technology and Design (Singapore) in 2014-2015, the University of Texas at Austin (USA) in 2013, Supelec (France) in 2012, and Alcatel-Lucent (Italy) in 2009. He has co-authored 50+ IEEE publications with 1000+ citations, and is co-inventor of a dozen filed patent families on wireless communications and networking. He is deeply involved in the research community, serving as an editor for the \textsc{IEEE Transactions on Wireless Communications} and \textsc{IEEE Communications Letters}, and as a workshop or special session co-organizer at IEEE Globecom'17, Asilomar'18, and IEEE ICC'19. He is also a frequent speaker and his contributions include a workshop keynote at IEEE PIMRC '18, an industry seminar at IEEE ICC'19, and tutorials at IEEE WCNC'18, IEEE ICC'18, IEEE Globecom'18, and IEEE PIMRC'19. He is the recipient of the IEEE ComSoc Outstanding Young Researcher Award for Europe, Middle-East \& Africa 2018.
\end{IEEEbiography}\vfill

\begin{IEEEbiography}[{\includegraphics[width=1in,height=1.25in,keepaspectratio]{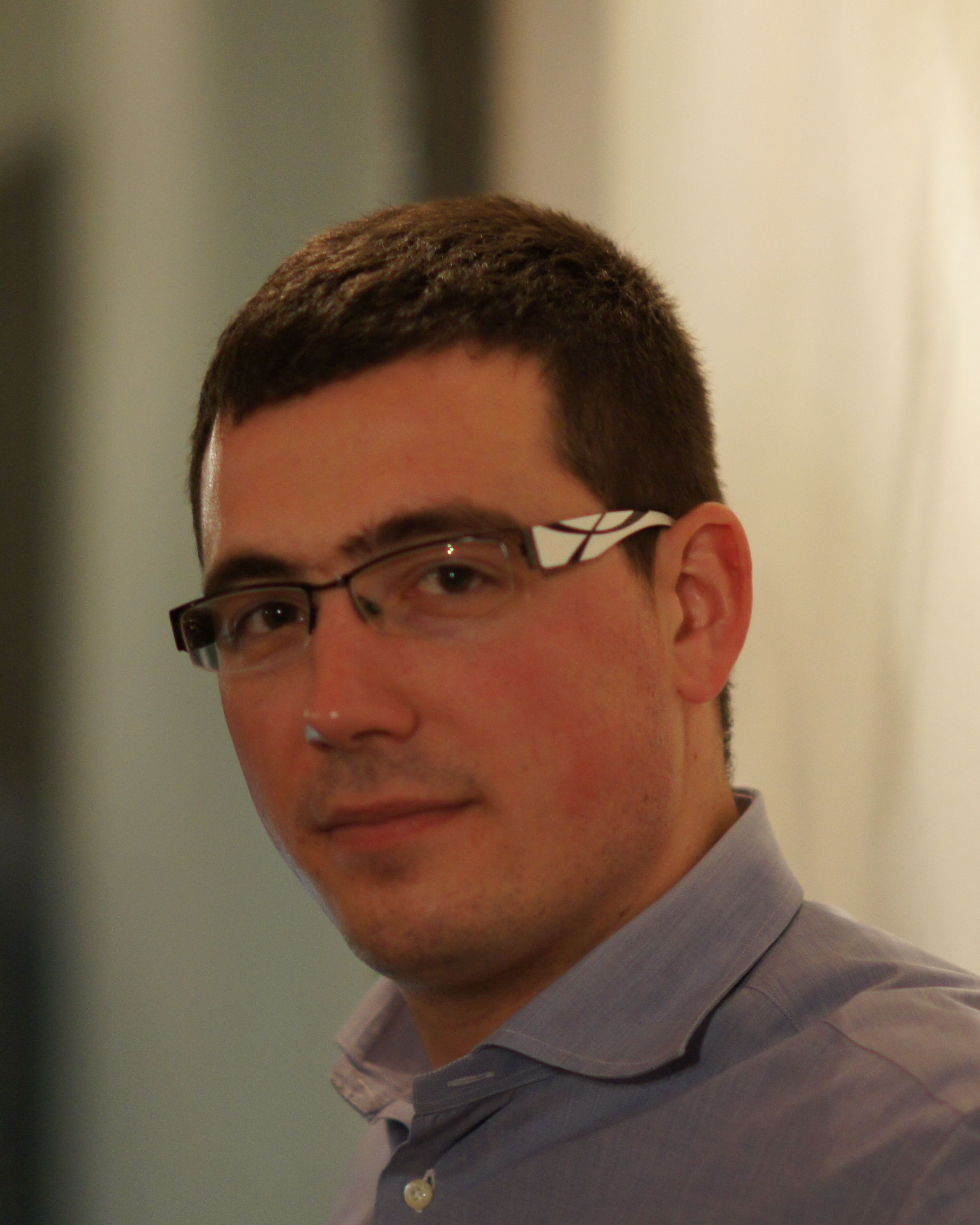}}]%
    {Paolo Baracca} (S'11--M'13) is a Research Engineer at Nokia Bell Labs in Stuttgart, Germany, since 2013. He received the B.Sc. and the M.Sc. degree in Telecommunications Engineering in 2007 and 2009, respectively, and the Ph.D. degree in Information Engineering in 2013, all from the University of Padova, Italy. His research interests include signal processing, multi-antenna techniques and scheduling for wireless communications. He has co-authored more than 30 research papers, holds more than 20 issued or pending patents, and regularly serves as reviewer of IEEE journals and conferences.
\end{IEEEbiography}

\begin{IEEEbiography}[{\includegraphics[width=1in,height=1.25in,keepaspectratio]{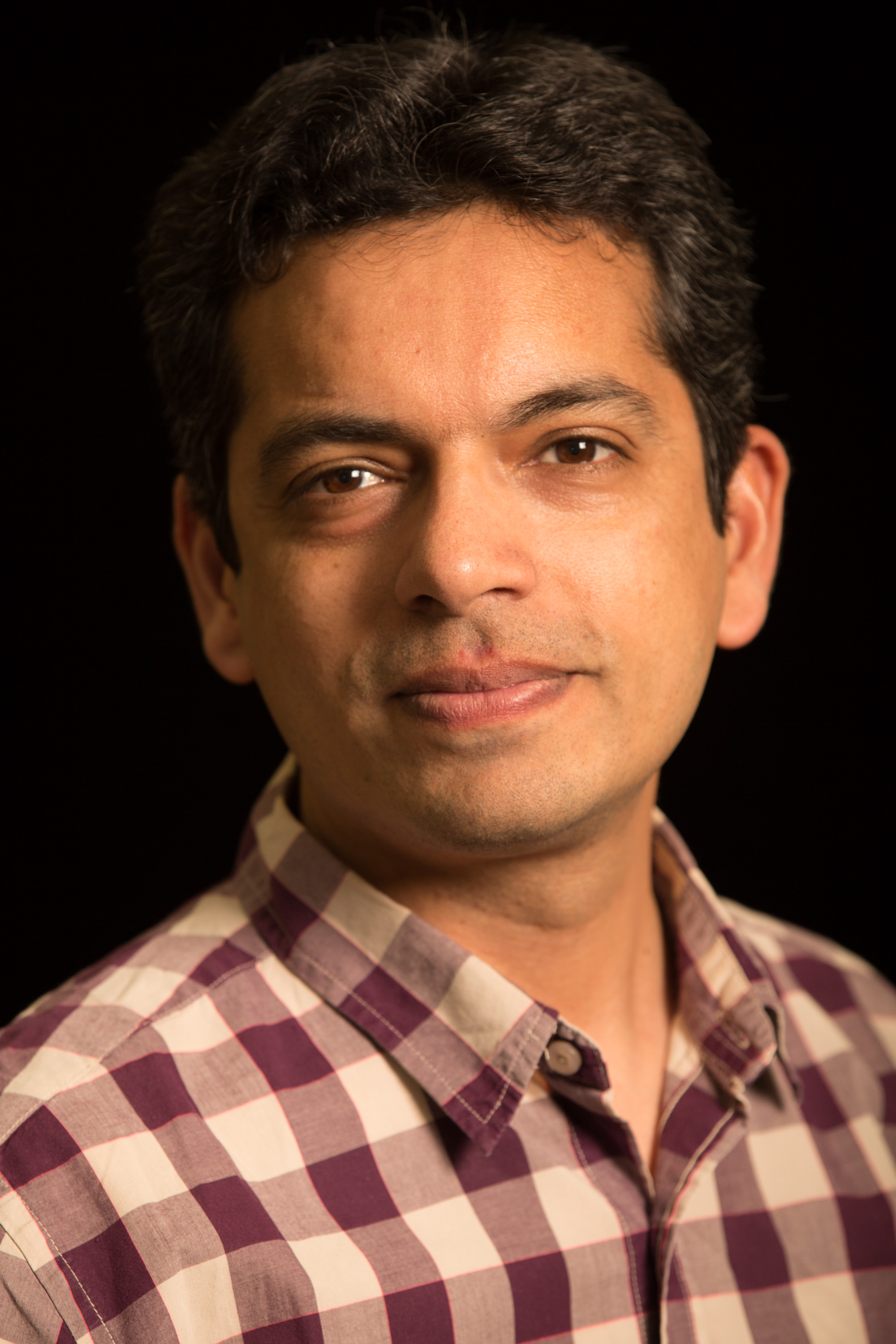}}]%
	{M. Majid Butt} (S'07--M'10--SM'15) received the M.Sc. degree in digital communications from Christian Albrechts University, Kiel, Germany, in 2005, and the Ph.D. degree in telecommunications from the Norwegian University of Science and Technology, Trondheim, Norway, in 2011. He is a Senior Scientist 5G+ Research at Nokia Bell Labs, Paris-Saclay, France, and a Visiting Research Assistant Professor at Trinity College Dublin, Dublin, Ireland. Prior to that, he has held various positions at the University of Glasgow, Glasgow, U.K., Trinity College Dublin, Fraunhofer HHI, Berlin, Germany, and the University of Luxembourg, Luxembourg City, Luxembourg. His current research interests include communication techniques for wireless networks with a focus on radio resource allocation, scheduling algorithms, energy efficiency, and machine learning for RAN. He has authored more than 60 peer-reviewed conference and journal publications in these areas. Dr. Butt was a recipient of the Marie Curie Alain Bensoussan Post-Doctoral Fellowship from the European Research Consortium for Informatics and Mathematics (ERCIM). He has served as the Organizer/Chair for various technical workshops on various aspects of communication systems in conjunction with major IEEE conferences, including Wireless Communications and Networking Conference, Globecom, and Greencom. He has been an Associate Editor for the \textsc{IEEE Access} and the \textsc{IEEE Communication Magazine} since 2016.
\end{IEEEbiography}

\begin{IEEEbiography}[{\includegraphics[width=1in,height=1.25in,keepaspectratio]{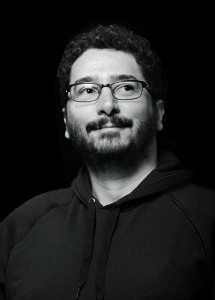}}]%
	{Nicola Marchetti} (M'13--SM'15) is currently Assistant Professor in Wireless Communications at Trinity College Dublin, Ireland. 
	He performs his research under the Irish Research Centre for Future Networks and Communications (CONNECT), 
	where he leads the Wireless Engineering and Complexity Science (WhyCOM) lab. 
	He received the Ph.D. in Wireless Communications from Aalborg University, Denmark in 2007, 
	and the M.Sc. in Electronic Engineering from University of Ferrara, Italy in 2003. 
	He also holds an M.Sc. in Mathematics which he received from Aalborg University in 2010. 
	His collaborations include research projects in cooperation with Nokia Bell Labs 
	and US Air Force Office of Scientific Research, among others. His research interests include Radio Resource Management, 
	Self-Organising Networks, Complex Systems Science and Signal Processing for communication networks. 
	He has authored in excess of 120 journals and conference papers, 2 books and 8 book chapters, holds 3 patents, 
	and received 4 best paper awards. He is a senior member of IEEE and serves as an associate editor for the \textsc{IEEE Internet of 
	Things Journal} since 2018 and the \textsc{EURASIP Journal on Wireless Communications and Networking} since 2017.
\end{IEEEbiography}\vfill

\end{document}